\documentclass[nonacm]{acmart}

\usepackage{todonotes}      
\usepackage{enumitem}       
\usepackage{array,ragged2e}                 
\newcolumntype{L}[1]{>{\RaggedRight}p{#1}}  
\newcolumntype{R}[1]{>{\RaggedLeft}p{#1}}   
\newcolumntype{C}[1]{>{\Centering}p{#1}}    
\newcolumntype{M}[1]{>{\RaggedRight}m{#1}}  
\newcolumntype{B}[1]{>{\RaggedRight}b{#1}}  
\usepackage{soul}           

\begin{document}

\acmBooktitle{arXiv}

\title{We need to aim at the top: Factors associated with cybersecurity awareness of cyber and information security decision-makers}
\renewcommand{\shorttitle}{We need to aim at the top}

\author{Simon Vrhovec}
\email{simon.vrhovec@um.si}
\affiliation{
  \institution{University of Maribor}
  \city{Maribor}
  \country{Slovenia}
}
\author{Blaž Markelj}
\email{blaz.markelj@um.si}
\affiliation{
  \institution{University of Maribor}
  \city{Maribor}
  \country{Slovenia}
}
\renewcommand{\shortauthors}{Vrhovec and Markelj}

\begin{abstract}
Cyberattacks pose a significant business risk to organizations. Although there is ample literature focusing on why people pose a major risk to organizational cybersecurity and how to deal with it, there is surprisingly little we know about cyber and information security decision-makers who are essentially the people in charge of setting up and maintaining organizational cybersecurity. In this paper, we study cybersecurity awareness of cyber and information security decision-makers, and investigate factors associated with it. We conducted an online survey among Slovenian cyber and information security decision-makers ($N=283$) to (1) determine whether their cybersecurity awareness is associated with adoption of antimalware solutions in their organizations, and (2) explore which organizational factors and personal characteristics are associated with their cybersecurity awareness. Our findings indicate that awareness of well-known threats and solutions seems to be quite low for individuals in decision-making roles. They also provide insights into which threats (e.g., DDoS attacks, botnets, industrial espionage, and phishing) and solutions (e.g., security operation center (SOC), advanced antimalware solutions with EDR/XDR capabilities, organizational critical infrastructure access control, centralized device management, multi-factor authentication, centralized management of software updates, and remote data deletion on lost or stolen devices) are cyber and information security decision-makers the least aware of. We uncovered that awareness of certain threats and solutions is positively associated with either adoption of advanced antimalware solutions with EDR/XDR capabilities or adoption of SOC. Additionally, we identified significant organizational factors (organizational role type) and personal characteristics (gender, age, experience with information security and experience with IT) related to cybersecurity awareness of cyber and information security decision-makers. Organization size and formal education were not significant. These results offer insights that can be leveraged in targeted cybersecurity training tailored to the needs of groups of cyber and information security decision-makers based on these key factors.
\end{abstract}


\maketitle

\section{Introduction}


The multitude of cyberattacks and increasing cybercrime is increasingly leading organizations to accept cybersecurity as a significant business risk that can cause major financial loses, damage to reputation and legal liabilities \cite{Batrachenko2024,Naseer2023,Prebot2023,Yin2023}. Decision-makers who are tasked with managing these cyber-risks need to deal with the challenges of evaluating risks in an evolving threat landscape and with constantly emerging new technologies \cite{Parkin2023,Roman2023}. It is already challenging how to determine the real causes of cyber-incidents \cite{Ebert2023} which makes identifying the right countermeasures before incidents happen, taking into account their cost-benefit relationships, an exceedingly difficult task \cite{Kianpour2021,Liu2022}. Although a systematic approach to cybersecurity seems imperial, there are ideas about less measuring cybersecurity and more communicating about it \cite{ThorntonTrump2023}. For example, organizational leaders, such as Chief Information Officer (CIO), Chief Security Information Officer (CISO), and Chief Technology Officer (CTO), are at the core of supporting cybersecurity strategies by improving governance and integration as well as fostering a new cultural mindset for cyber-resiliency \cite{Loonam2020}. Nevertheless, decision-makers are primarily tasked with decisions on cybersecurity measures that can range from adopting cybersecurity standards, technical measures, such as advanced antimalware \cite{Smmarwar2024} and intrusion detection \cite{Song2024} solutions, and human-centric measures, such as various types of cybersecurity training \cite{Fujs2023:cs,Tian2023}, to inter-organizational measures, such as cyberthreat intelligence \cite{Preuveneers2023,Turner2023,Dykstra2023,Piazza2023}.

The human aspect of cybersecurity is a thriving research area. Studies focus on social engineering and phishing
\cite{Cuchta2023},
decision-making processes of SOC analysts \cite{Reeves2023}, security concerns associated with adoption of social robots \cite{Zvanut2024}, effects of cyberattack proximity \cite{Gomez2024}, fake and real news decision-making \cite{Brockinton2022}, information seeking \cite{Vrhovec2023:cs}, testing cyber soldiers \cite{Lif2022}, replacing aging and thus insecure smart devices \cite{Lenz2023:cs}, etc. These are just a few examples which paint the diversity of these studies. The published literature however seldom focuses on decision-makers even though they are among the key enablers of cybersecurity in organizations. For example, Bongiovanni et al. (2022) \cite{Bongiovanni2022} investigated how decision-makers implement measures recommended by published cybersecurity guidelines. A study on adoption of cybersecurity standards in SMEs found significant factors, such as demographics (organization size, intensity of IT usage, number of IT staff, number of IT security staff, investments in IT, investments in IT security), attitudes towards organizational cybersecurity risks and customer cybersecurity needs \cite{Auyporn2023}. Triplett (2022) \cite{Triplett2022} studied how cybersecurity is promoted by organizational leaders. Studies indicate that unintentional human factors which facilitate cyberattacks, include lack of support by leaders, lack of knowledge and skills, being not aware of severity and damage cyberattacks can have, complacency and naivety coupled with reluctance to learn or seek help, and cybersecurity fatigue \cite{Mikuletic2024}. Some studies indicate a lack of awareness and cybersecurity education by decision-makers \cite{Drape2021,Rawindaran2022}. It is especially worrisome that decision-makers are often unaware of what solutions can do to protect data \cite{Rawindaran2022}. Formal education and organization size were found to be contributing factors \cite{Rawindaran2022}. Nevertheless, we found a single study that suggests cyberthreat awareness among decision-makers is high \cite{Moyo2021} albeit it remains unclear how the researchers came to this conclusion since they did not report it in their work. The findings found in the literature are thus mixed at best. The literature also does not provide any insights into which factors may be associated with cybersecurity awareness of decision-makers. These insights may would to improve the awareness of cyber and information security decision-makers, e.g., with targeted cybersecurity training tailored to the needs of groups of decision-makers based on these key factors.

In this paper, we focus on cybersecurity awareness of decision-makers by investigating their awareness of well-known threats and solutions. First, we investigate whether there are differences between cybersecurity awareness of decision-makers in organizations adopting advanced antimalware solutions, and in organizations that do not. We break down advanced antimalware solutions into (1) advanced antimalware solutions with EDR/XDR capabilities, and (2) security operation center (SOC) as two of the most common advanced antimalware solutions found on the market today. This enabled us to determine whether cybersecurity awareness might play an important role in their decisions on cybersecurity. Second, we explored which organizational factors and personal characteristics are associated with cybersecurity awareness of decision-makers. Based on this, we developed these research questions:

\begin{itemize}[leftmargin=1.4cm]
    \item[\textit{RQ1}:]{Are there differences in cybersecurity awareness of decision-makers across groups based on adoption of antimalware solutions in their organizations?}
    \item[\textit{RQ2a}:]{Are there differences in cybersecurity awareness of decision-makers across groups based on organizational factors?}
    \item[\textit{RQ2b}:]{Are there differences in cybersecurity awareness of decision-makers across groups based on their personal characteristics?}
\end{itemize}

This paper is structured as follows. We present the research methodology in Section~\ref{section:method}. The results of our study are presented in Section~\ref{section:results}. We provide theoretical and practical implications as well as limitations and directions for future work in Section~\ref{section:discussion}.

\section{Methodology}
\label{section:method}

\subsection{Research design}

We employed a cross-sectional research design to capture cybersecurity awareness of decision-makers and other variables at a specific time period. We conducted a survey among cyber and information security decision-makers in Slovenian organizations through the CINT platform. We included IT/IS executive (e.g., CISO, CIO), non-IT/IS executive (e.g., CEO, CFO) and non-executive (e.g., IT administrator, department head) decision-makers in the sample to cover the whole spectrum of cyber and information security decision-makers in organizations of different sizes, and enable comparison between them.

\subsection{Ethical considerations}

This study involved human participants. The study proposal was approved on 27 February 2023 by the Research Ethics Committee of the Faculty of Criminal Justice and Security at the University of Maribor [2702-2023].

\subsection{Measurement instrument}

The survey questionnaire was designed to measure \textit{awareness of threats}, \textit{awareness of solutions}, \textit{adopted antimalware solution type} and \textit{adopted SOC realization} in addition to organizational factors (\textit{organizational role type} and \textit{organization size}) and personal characteristics (\textit{gender}, \textit{formal education}, \textit{age}, \textit{experience with information security} and \textit{experience with IT}).

\textit{Awareness of threats} included questions on nine common threats: loss of access to data (e.g., ransomware, locking of devices); information system intrusion (e.g., hacking); theft of business-critical data (industrial espionage); distributed denial of service (DDoS) attacks; takeover of devices (e.g., botnets); malware infection (e.g., viruses, worms, Trojans, spyware); phishing; online fraud (e.g., CEO fraud, business email compromise); and internal threats (e.g., deliberate data deletion, unauthorized data access, unauthorized personal devices). \textit{Awareness of solutions} included questions on 12 available solutions for managing cyber-risks: remote data deletion on lost or stolen devices; advanced antimalware solutions (with EDR/XDR capabilities); secure connection (e.g., VPN); cloud synchronization of data; data backup; centralized device management, including mobile device management (MDM); advanced firewalls (with IPS/IDS capabilities); training on secure use of devices; multi-factor authentication (e.g., 2FA); security operation center (SOC) 24/7; centralized management of software updates; and organizational critical infrastructure access control. \textit{Awareness of threats} and \textit{awareness of solutions} were measured on 4-point ordinal scales adapted from \cite{Cochran2021}: 1 -- "I am very familiar with this [threat/solution] and what it is"; 2 -- "I have heard about this [threat/solution] previously and am somewhat familiar with it"; 3 -- "I have heard mention of this [threat/solution] before but am largely unfamiliar with it"; 4 -- "I was not aware of this [threat/solution] before today".

Respondents were first shown a description of advanced antimalware solutions, including descriptions of EDR and XDR capabilities as well as SOC. Then, respondents were asked which types of antimalware solutions and SOC realizations were adopted in their respective organizations. Options for \textit{adopted antimalware solution type} included an advanced antimalware solution with EDR/XDR capabilities, a standard antimalware solution, such as antivirus programs, and none. Options for \textit{adopted SOC realization} included a dedicated 24/7 security team within a respondent's organization, an external SOC (SOC-as-a-service model), and none (incidents handled by IT or information security department).

\subsection{Sample and data collection}

The survey was conducted in March 2023 through the CINT platform. We used organizational role as a screening question to ensure that all respondents were qualified to take the survey. A total of $356$ respondents took the survey. After excluding unqualified respondents ($21$), responses with over 10 percent missing values ($5$), and responses that indicated non-engagement bias ($47$), we were left with $N=283$ useful responses. Table~\ref{table:sample} presents the key sample characteristics.

\begin{table}[!htb]
\footnotesize 
\caption{\label{table:sample} Sample characteristics.}
\begin{tabular}{llrr}
\toprule
Characteristic &  & Frequency & Percent \\ \midrule
Organization size & Micro & 133 & $47.0\%$ \\
 & Small & 88 & $31.1\%$ \\
 & Medium & 49 & $17.3\%$ \\
 & Large & 13 & $4.6\%$ \\
Role & CEO & 78 & $27.6\%$ \\
 & CFO & 36 & $12.7\%$ \\
 & CIO & 50 & $17.7\%$ \\
 & IT executive & 33 & $11.7\%$ \\
 & CISO & 22 & $7.8\%$ \\
 & Other & 64 & $22.6\%$ \\
Gender & Female & 87 & $30.7\%$ \\
 & Male & 195 & $68.9\%$ \\
 & N/A & 1 & $0.4\%$ \\
Formal education & Finished high school & 54 & $19.1\%$ \\
 & Bachelor’s degree & 90 & $31.8\%$ \\
 & Master’s degree & 106 & $37.5\%$ \\
 & PhD degree & 33 & $11.7\%$ \\
\bottomrule
\end{tabular}
\end{table}

Average respondent \textit{age} was $36.7$ years ($SD=12.1$). Mean \textit{experience with IT} was $10.4$ years ($SD=8.5$), and mean \textit{experience with information security} was $9.9$ years ($SD=8.1$). Average \textit{employment duration} was $10.4$ years ($SD=9.1$).

\subsection{Data analysis}

We used R version 4.3.3 with packages \textit{psych} version 2.3.9, \textit{dplyr} version 1.1.3, \textit{FSA} version 0.9.5 and \textit{car} version 3.1-2 for data analysis. We merged some categories and scores before data analysis. First, we merged categories for \textit{organizational role} into categories IT/IS executive (e.g., CISO, CIO), non-IT/IS executive (e.g., CEO, CFO) and non-executive (e.g., IT administrator, department head). This enabled us to test whether the \textit{organizational role type} according to IT/IS background and/or position within an organization is associated with cybersecurity awareness of decision-makers. Next, we merged scores for \textit{awareness of threats} and \textit{awareness of solutions} to improve the reliability of measuring these constructs. Measuring awareness with self-reported questionnaires is challenging as people do not tend to be very realistic about their own awareness of specific phenomena. We used a scale that adequately addresses this issue by describing four clearly distinct levels of awareness. We decided to further improve the reliability of the used scale by dividing respondents into those that are at least somewhat familiar with a threat/solution (\textit{aware}) and those who are largely unfamiliar with it or heard about it for the first time during the survey (\textit{not aware}). A beneficial side-effect of this mergence is an improved interpretability of the study results enabling the comparison of the awareness level of respondents in different groups.

Data analysis included non-parametric tests, such as Kruskal-Wallis test for determining differences between three or more groups with post-hoc Dunn's tests with Bonferroni correction for determining differences between pairs of those groups. We used Wilcoxon tests to determine differences between two groups. We also used parametric tests when possible -- we conducted independent samples $t$ tests for determining differences in means across two groups. We used standard significance levels ($\alpha = 0.05$, $\alpha = 0.01$, and $\alpha = 0.001$) for determining significance of all statistical test results.

\section{Results}
\label{section:results}

In this section, we present the results relevant for answering the posed research questions. We analyzed the data related to cybersecurity awareness of decision-makers separately for all research questions.

\subsection{RQ1: Are there differences in cybersecurity awareness of decision-makers across groups based on adoption of antimalware solutions in their organizations?}

We analyzed cybersecurity awareness of cyber and information security decision-makers separately for two of the most common advanced antimalware solutions found on the market today: (1) advanced antimalware solutions with EDR/XDR capabilities, and (2) security operation centers (SOC). Since the former is a technical solution, we compared it with other types of technical solutions (i.e., a standard antimalware solution and none). The latter is an organizational solution which can have varying realizations (i.e., internal SOC, external SOC, none).

\subsubsection{Adopted antimalware solution type}

Table~\ref{table:am-t} presents the differences in respondents' \textit{awareness of threats} for groups based on their organization's \textit{adopted antimalware solution type}. The results indicate significant differences for awareness of six threats. However, there were only four threats with clearly distinguishable differences between different types of adopted antimalware solutions.
For three of these threats (i.e., industrial espionage, botnets and phishing), there were significant differences between organizations that have either antimalware solution (EDR/XDR or standard) and those which do not. In all cases, respondents in the latter had a significantly lower awareness of these threats.
Awareness of DDoS attacks was however significantly higher for respondents in organizations adopting advanced antimalware solutions with EDR/XDR capabilities than respondents in those adopting a standard antimalware solution or not adopting any at all.

\begin{table}[!htb]
\footnotesize 
\caption{\label{table:am-t} Differences in awareness of threats across adopted antimalware solution types.}
\begin{tabular}{lrrrrlrrrr}
\toprule
 & 1: EDR/XDR & 2: Standard & 3: None & $H$ &  & $p$ & $p_{1-2}$ & $p_{1-3}$ & $p_{2-3}$ \\ \midrule
\multicolumn{10}{l}{\textit{Loss of access to data (e.g., ransomware, locking of devices)}} \\
Not aware & $10 (13.0\%)$ & $26 (19.0\%)$ & $14 (33.3\%)$ & $7.032$ & $^{*}$ & $0.0297$ & $0.9190$ & $0.0250$ & $0.1223$ \\
Aware & $66 (85.7\%)$ & $111 (81.0\%)$ & $28 (66.7\%)$ &  &  &  &  &  &  \\
\multicolumn{10}{l}{\textit{Information system intrusion (e.g., hacking)}} \\
Not aware & $14 (18.2\%)$ & $18 (13.1\%)$ & $8 (19.0\%)$ & $1.392$ &  & $0.4986$ &  &  &  \\
Aware & $63 (81.8\%)$ & $119 (86.9\%)$ & $34 (81.0\%)$ &  &  &  &  &  &  \\
\multicolumn{10}{l}{\textit{Theft of business-critical data (industrial espionage)}} \\
Not aware & $18 (23.4\%)$ & $34 (24.8\%)$ & $20 (47.6\%)$ & $9.459$ & $^{**}$ & $0.0088$ & $1.0000$ & $0.0151$ & $0.0123$ \\
Aware & $59 (76.6\%)$ & $103 (75.2\%)$ & $22 (52.4\%)$ &  &  &  &  &  &  \\
\multicolumn{10}{l}{\textit{Distributed denial of service (DDoS) attacks}} \\
Not aware & $11 (14.3\%)$ & $56 (40.9\%)$ & $21 (50.0\%)$ & $20.335$ & $^{***}$ & $0.0000$ & $0.0003$ & $0.0003$ & $0.8324$ \\
Aware & $65 (84.4\%)$ & $81 (59.1\%)$ & $21 (50.0\%)$ &  &  &  &  &  &  \\
\multicolumn{10}{l}{\textit{Takeover of devices (e.g., botnets)}} \\
Not aware & $15 (19.5\%)$ & $41 (29.9\%)$ & $22 (52.4\%)$ & $13.537$ & $^{**}$ & $0.0011$ & $0.3474$ & $0.0007$ & $0.0193$ \\
Aware & $61 (79.2\%)$ & $95 (69.3\%)$ & $20 (47.6\%)$ &  &  &  &  &  &  \\
\multicolumn{10}{l}{\textit{Malware infection (e.g., viruses, worms, Trojans, spyware)}} \\
Not aware & $9 (11.7\%)$ & $19 (13.9\%)$ & $6 (14.3\%)$ & $0.180$ &  & $0.9137$ &  &  &  \\
Aware & $66 (85.7\%)$ & $118 (86.1\%)$ & $36 (85.7\%)$ &  &  &  &  &  &  \\
\multicolumn{10}{l}{\textit{Phishing}} \\
Not aware & $11 (14.3\%)$ & $31 (22.6\%)$ & $18 (42.9\%)$ & $12.422$ & $^{**}$ & $0.0020$ & $0.5028$ & $0.0013$ & $0.0207$ \\
Aware & $66 (85.7\%)$ & $106 (77.4\%)$ & $24 (57.1\%)$ &  &  &  &  &  &  \\
\multicolumn{10}{l}{\textit{Online fraud (e.g., CEO fraud, business email compromise)}} \\
Not aware & $15 (19.5\%)$ & $30 (21.9\%)$ & $10 (23.8\%)$ & $0.330$ &  & $0.8477$ &  &  &  \\
Aware & $62 (80.5\%)$ & $107 (78.1\%)$ & $32 (76.2\%)$ &  &  &  &  &  &  \\
\multicolumn{10}{l}{\textit{Internal threats (e.g., deliberate data deletion, unauthorized data access, unauthorized personal devices)}} \\
Not aware & $8 (10.4\%)$ & $30 (21.9\%)$ & $14 (33.3\%)$ & $9.263$ & $^{**}$ & $0.0097$ & $0.1445$ & $0.0087$ & $0.2989$ \\
Aware & $67 (87.0\%)$ & $105 (76.6\%)$ & $27 (64.3\%)$ &  &  &  &  &  &  \\
\bottomrule
\end{tabular}
\begin{flushleft}
We conducted Kruskal-Wallis tests to determine whether there are differences in awareness of threats across groups. For threats with significant differences across groups, we conducted post-hoc Dunn's tests with Bonferroni correction to determine which groups significantly differ from each other. Notes: $^{*} p < 0.05, ^{**} p < 0.01, ^{***} p < 0.001$.
\end{flushleft}
\end{table}

Table~\ref{table:am-s} presents the differences in respondents' \textit{awareness of solutions} for groups based on their organization's \textit{adopted antimalware solution type}. The results indicate significant differences for awareness of 10 solutions. There were however just eight clear differences between different types of adopted antimalware solutions.
For six of these solutions (i.e., remote deletion, advanced firewalls, training, multi-factor authentication, SOC and centralized management of software updates), there were significant differences between respondents in organizations that have either antimalware solution (EDR/XDR or standard) and respondents in those which do not. In all cases, respondents in the latter had significantly lower awareness of these solutions.
For two solutions (i.e., advanced antimalware solutions with EDR/XDR capabilities and centralized device management), there were clear differences among all three adopted antimalware solution type groups. In both cases, the most aware were respondents in organizations adopting advanced antimalware solutions with EDR/XDR capabilities, followed by respondents in those which adopted a standard antimalware solution, while the least aware were respondents in organizations that did not adopt any antimalware solution.

\begin{table}[!htb]
\footnotesize 
\caption{\label{table:am-s} Differences in awareness of solutions across adopted antimalware solution types.}
\begin{tabular}{lrrrrlrrrr}
\toprule
 & 1: EDR/XDR & 2: Standard & 3: None & $H$ &  & $p$ & $p_{1-2}$ & $p_{1-3}$ & $p_{2-3}$ \\ \midrule
\multicolumn{10}{l}{\textit{Remote data deletion on lost or stolen devices}} \\
Not aware & $12 (15.6\%)$ & $34 (24.8\%)$ & $19 (45.2\%)$ & $12.081$ & $^{**}$ & $0.0024$ & $0.4587$ & $0.0016$ & $0.0265$ \\
Aware & $63 (81.8\%)$ & $102 (74.5\%)$ & $23 (54.8\%)$ &  &  &  &  &  &  \\
\multicolumn{10}{l}{\textit{Advanced antimalware solutions (with EDR/XDR capabilities)}} \\
Not aware & $9 (11.7\%)$ & $53 (38.7\%)$ & $27 (64.3\%)$ & $35.845$ & $^{***}$ & $0.0000$ & $0.0002$ & $0.0000$ & $0.0055$ \\
Aware & $67 (87.0\%)$ & $82 (59.9\%)$ & $14 (33.3\%)$ &  &  &  &  &  &  \\
\multicolumn{10}{l}{\textit{Secure connection (e.g., VPN)}} \\
Not aware & $15 (19.5\%)$ & $23 (16.8\%)$ & $6 (14.3\%)$ & $0.546$ &  & $0.7611$ &  &  &  \\
Aware & $62 (80.5\%)$ & $114 (83.2\%)$ & $36 (85.7\%)$ &  &  &  &  &  &  \\
\multicolumn{10}{l}{\textit{Cloud synchronization of data}} \\
Not aware & $15 (19.5\%)$ & $16 (11.7\%)$ & $13 (31.0\%)$ & $8.624$ & $^{*}$ & $0.0134$ & $0.4590$ & $0.3426$ & $0.0123$ \\
Aware & $62 (80.5\%)$ & $120 (87.6\%)$ & $29 (69.0\%)$ &  &  &  &  &  &  \\
\multicolumn{10}{l}{\textit{Data backup}} \\
Not aware & $13 (16.9\%)$ & $20 (14.6\%)$ & $9 (21.4\%)$ & $1.072$ &  & $0.5850$ &  &  &  \\
Aware & $63 (81.8\%)$ & $116 (84.7\%)$ & $33 (78.6\%)$ &  &  &  &  &  &  \\
\multicolumn{10}{l}{\textit{Centralized device management, including mobile device management (MDM)}} \\
Not aware & $7 (9.1\%)$ & $38 (27.7\%)$ & $22 (52.4\%)$ & $26.585$ & $^{***}$ & $0.0000$ & $0.0075$ & $0.0000$ & $0.0057$ \\
Aware & $70 (90.9\%)$ & $97 (70.8\%)$ & $20 (47.6\%)$ &  &  &  &  &  &  \\
\multicolumn{10}{l}{\textit{Advanced firewalls (with IPS/IDS capabilities)}} \\
Not aware & $12 (15.6\%)$ & $24 (17.5\%)$ & $18 (42.9\%)$ & $14.229$ & $^{***}$ & $0.0008$ & $1.0000$ & $0.0015$ & $0.0015$ \\
Aware & $65 (84.4\%)$ & $112 (81.8\%)$ & $24 (57.1\%)$ &  &  &  &  &  &  \\
\multicolumn{10}{l}{\textit{Training on secure use of devices}} \\
Not aware & $9 (11.7\%)$ & $28 (20.4\%)$ & $16 (38.1\%)$ & $11.260$ & $^{**}$ & $0.0036$ & $0.4009$ & $0.0024$ & $0.0446$ \\
Aware & $67 (87.0\%)$ & $108 (78.8\%)$ & $26 (61.9\%)$ &  &  &  &  &  &  \\
\multicolumn{10}{l}{\textit{Multi-factor authentication (e.g., 2FA)}} \\
Not aware & $12 (15.6\%)$ & $30 (21.9\%)$ & $27 (64.3\%)$ & $36.207$ & $^{***}$ & $0.0000$ & $0.9234$ & $0.0000$ & $0.0000$ \\
Aware & $65 (84.4\%)$ & $106 (77.4\%)$ & $15 (35.7\%)$ &  &  &  &  &  &  \\
\multicolumn{10}{l}{\textit{Security operation center (SOC) 24/7}} \\
Not aware & $16 (20.8\%)$ & $52 (38.0\%)$ & $28 (66.7\%)$ & $23.445$ & $^{***}$ & $0.0000$ & $0.0517$ & $0.0000$ & $0.0024$ \\
Aware & $59 (76.6\%)$ & $85 (62.0\%)$ & $14 (33.3\%)$ &  &  &  &  &  &  \\
\multicolumn{10}{l}{\textit{Centralized management of software updates}} \\
Not aware & $11 (14.3\%)$ & $35 (25.5\%)$ & $22 (52.4\%)$ & $21.232$ & $^{***}$ & $0.0000$ & $0.2111$ & $0.0000$ & $0.0012$ \\
Aware & $66 (85.7\%)$ & $101 (73.7\%)$ & $19 (45.2\%)$ &  &  &  &  &  &  \\
\multicolumn{10}{l}{\textit{Organizational critical infrastructure access control}} \\
Not aware & $13 (16.9\%)$ & $38 (27.7\%)$ & $19 (45.2\%)$ & $10.445$ & $^{**}$ & $0.0054$ & $0.2997$ & $0.0037$ & $0.0865$ \\
Aware & $62 (80.5\%)$ & $98 (71.5\%)$ & $23 (54.8\%)$ &  &  &  &  &  &  \\
\bottomrule
\end{tabular}
\begin{flushleft}
We conducted Kruskal-Wallis tests to determine whether there are differences in awareness of solutions across groups. For solutions with significant differences across groups, we conducted post-hoc Dunn's tests with Bonferroni correction to determine which groups significantly differ from each other. Notes: $^{*} p < 0.05, ^{**} p < 0.01, ^{***} p < 0.001$.
\end{flushleft}
\end{table}

These results indicate that awareness of certain threats and solutions is positively associated with adoption of antimalware solutions.
Decision-makers in organizations adopting an advanced antimalware solution with EDR/XDR capabilities are more aware of one threat (i.e., DDoS attacks) and two solutions (i.e., advanced antimalware solutions with EDR/XDR capabilities and centralized device management) than decision-makers in organizations adopting a standard antimalware solution. Additionally, they are more aware of four threats (i.e., industrial espionage, botnets, phishing and DDoS attacks) and eight solutions (i.e., remote deletion, advanced firewalls, training, multi-factor authentication, SOC, centralized management of software updates, advanced antimalware solutions with EDR/XDR capabilities and centralized device management) than decision-makers in organizations which do not adopt any antimalware solution.
Decision-makers in organizations adopting a standard antimalware solution are similarly more aware of three threats (i.e., industrial espionage, botnets and phishing) and eight solutions (i.e., remote deletion, advanced firewalls, training, multi-factor authentication, SOC, centralized management of software updates, advanced antimalware solutions with EDR/XDR capabilities and centralized device management) than decision-makers in organizations which do not adopt any antimalware solution.

\subsubsection{Adopted SOC realization}

Table~\ref{table:soc-t} presents the differences in respondents' \textit{awareness of threats} for groups based on \textit{adopted SOC realization} in their organizations. The results indicate significant differences in awareness of five threats.
A single threat (i.e., online fraud) had clearly distinguishable differences between different realizations of SOC. Respondents in organizations adopting an internal SOC were markedly more aware of this threat than respondents in organizations adopting an external SOC or not having a SOC.

\begin{table}[!htb]
\footnotesize 
\caption{\label{table:soc-t} Differences in awareness of threats across adopted SOC realizations.}
\begin{tabular}{lrrrrlrrrr}
\toprule
 & 1: Internal SOC & 2: External SOC & 3: None & $H$ &  & $p$ & $p_{1-2}$ & $p_{1-3}$ & $p_{2-3}$ \\ \midrule
\multicolumn{10}{l}{\textit{Loss of access to data (e.g., ransomware, locking of devices)}} \\
Not aware & $5 (7.5\%)$ & $11 (17.7\%)$ & $30 (25.4\%)$ & $8.883$ & $^{*}$ & $0.0118$ & $0.4237$ & $0.0089$ & $0.6301$ \\
Aware & $61 (91.0\%)$ & $51 (82.3\%)$ & $88 (74.6\%)$ &  &  &  &  &  &  \\
\multicolumn{10}{l}{\textit{Information system intrusion (e.g., hacking)}} \\
Not aware & $7 (10.4\%)$ & $11 (17.7\%)$ & $18 (15.3\%)$ & $1.454$ &  & $0.4834$ &  &  &  \\
Aware & $60 (89.6\%)$ & $51 (82.3\%)$ & $100 (84.7\%)$ &  &  &  &  &  &  \\
\multicolumn{10}{l}{\textit{Theft of business-critical data (industrial espionage)}} \\
Not aware & $14 (20.9\%)$ & $15 (24.2\%)$ & $34 (28.8\%)$ & $1.479$ &  & $0.4773$ &  &  &  \\
Aware & $53 (79.1\%)$ & $47 (75.8\%)$ & $84 (71.2\%)$ &  &  &  &  &  &  \\
\multicolumn{10}{l}{\textit{Distributed denial of service (DDoS) attacks}} \\
Not aware & $11 (16.4\%)$ & $21 (33.9\%)$ & $48 (40.7\%)$ & $11.144$ & $^{**}$ & $0.0038$ & $0.1147$ & $0.0026$ & $1.0000$ \\
Aware & $55 (82.1\%)$ & $41 (66.1\%)$ & $70 (59.3\%)$ &  &  &  &  &  &  \\
\multicolumn{10}{l}{\textit{Takeover of devices (e.g., botnets)}} \\
Not aware & $13 (19.4\%)$ & $21 (33.9\%)$ & $41 (34.7\%)$ & $5.072$ &  & $0.0792$ &  &  &  \\
Aware & $53 (79.1\%)$ & $41 (66.1\%)$ & $76 (64.4\%)$ &  &  &  &  &  &  \\
\multicolumn{10}{l}{\textit{Malware infection (e.g., viruses, worms, Trojans, spyware)}} \\
Not aware & $3 (4.5\%)$ & $14 (22.6\%)$ & $13 (11.0\%)$ & $9.904$ & $^{**}$ & $0.0071$ & $0.0057$ & $0.5824$ & $0.0787$ \\
Aware & $63 (94.0\%)$ & $48 (77.4\%)$ & $104 (88.1\%)$ &  &  &  &  &  &  \\
\multicolumn{10}{l}{\textit{Phishing}} \\
Not aware & $7 (10.4\%)$ & $20 (32.3\%)$ & $27 (22.9\%)$ & $9.067$ & $^{*}$ & $0.0107$ & $0.0084$ & $0.1491$ & $0.4467$ \\
Aware & $60 (89.6\%)$ & $42 (67.7\%)$ & $91 (77.1\%)$ &  &  &  &  &  &  \\
\multicolumn{10}{l}{\textit{Online fraud (e.g., CEO fraud, business email compromise)}} \\
Not aware & $3 (4.5\%)$ & $15 (24.2\%)$ & $33 (28.0\%)$ & $14.964$ & $^{***}$ & $0.0006$ & $0.0174$ & $0.0005$ & $1.0000$ \\
Aware & $64 (95.5\%)$ & $47 (75.8\%)$ & $85 (72.0\%)$ &  &  &  &  &  &  \\
\multicolumn{10}{l}{\textit{Internal threats (e.g., deliberate data deletion, unauthorized data access, unauthorized personal devices)}} \\
Not aware & $6 (9.0\%)$ & $10 (16.1\%)$ & $27 (22.9\%)$ & $5.759$ &  & $0.0562$ &  &  &  \\
Aware & $59 (88.1\%)$ & $52 (83.9\%)$ & $89 (75.4\%)$ &  &  &  &  &  &  \\
\bottomrule
\end{tabular}
\begin{flushleft}
We conducted Kruskal-Wallis tests to determine whether there are differences in awareness of threats across groups. For threats with significant differences across groups, we conducted post-hoc Dunn's tests with Bonferroni correction to determine which groups significantly differ from each other. Notes: $^{*} p < 0.05, ^{**} p < 0.01, ^{***} p < 0.001$.
\end{flushleft}
\end{table}

Table~\ref{table:soc-s} presents the differences in respondents' \textit{awareness of solutions} for groups based on \textit{adopted SOC realization} in their organizations. The results indicate significant differences in awareness of five solutions, with clear differences between different realizations of SOC for two of them.
For both SOC and critical infrastructure access control, there were significant differences between respondents in organizations that have either realization of SOC (internal or external) and respondents in those which do not. The latter had significantly lower awareness of both solutions.

\begin{table}[!htb]
\footnotesize 
\caption{\label{table:soc-s} Differences in awareness of solutions across adopted SOC realizations.}
\begin{tabular}{lrrrrlrrrr}
\toprule
 & 1: Internal SOC & 2: External SOC & 3: None & $H$ &  & $p$ & $p_{1-2}$ & $p_{1-3}$ & $p_{2-3}$ \\ \midrule
\multicolumn{10}{l}{\textit{Remote data deletion on lost or stolen devices}} \\
Not aware & $7 (10.4\%)$ & $13 (21.0\%)$ & $34 (28.8\%)$ & $8.144$ & $^{*}$ & $0.0171$ & $0.5018$ & $0.0134$ & $0.6467$ \\
Aware & $58 (86.6\%)$ & $49 (79.0\%)$ & $83 (70.3\%)$ &  &  &  &  &  &  \\
\multicolumn{10}{l}{\textit{Advanced antimalware solutions (with EDR/XDR capabilities)}} \\
Not aware & $12 (17.9\%)$ & $23 (37.1\%)$ & $45 (38.1\%)$ & $8.899$ & $^{*}$ & $0.0117$ & $0.0587$ & $0.0136$ & $1.0000$ \\
Aware & $54 (80.6\%)$ & $38 (61.3\%)$ & $71 (60.2\%)$ &  &  &  &  &  &  \\
\multicolumn{10}{l}{\textit{Secure connection (e.g., VPN)}} \\
Not aware & $11 (16.4\%)$ & $11 (17.7\%)$ & $20 (16.9\%)$ & $0.040$ &  & $0.9800$ &  &  &  \\
Aware & $56 (83.6\%)$ & $51 (82.3\%)$ & $98 (83.1\%)$ &  &  &  &  &  &  \\
\multicolumn{10}{l}{\textit{Cloud synchronization of data}} \\
Not aware & $9 (13.4\%)$ & $15 (24.2\%)$ & $18 (15.3\%)$ & $3.034$ &  & $0.2194$ &  &  &  \\
Aware & $57 (85.1\%)$ & $47 (75.8\%)$ & $100 (84.7\%)$ &  &  &  &  &  &  \\
\multicolumn{10}{l}{\textit{Data backup}} \\
Not aware & $9 (13.4\%)$ & $14 (22.6\%)$ & $19 (16.1\%)$ & $1.979$ &  & $0.3717$ &  &  &  \\
Aware & $58 (86.6\%)$ & $48 (77.4\%)$ & $97 (82.2\%)$ &  &  &  &  &  &  \\
\multicolumn{10}{l}{\textit{Centralized device management, including mobile device management (MDM)}} \\
Not aware & $8 (11.9\%)$ & $11 (17.7\%)$ & $34 (28.8\%)$ & $7.851$ & $^{*}$ & $0.0197$ & $1.0000$ & $0.0230$ & $0.2423$ \\
Aware & $58 (86.6\%)$ & $51 (82.3\%)$ & $83 (70.3\%)$ &  &  &  &  &  &  \\
\multicolumn{10}{l}{\textit{Advanced firewalls (with IPS/IDS capabilities)}} \\
Not aware & $10 (14.9\%)$ & $10 (16.1\%)$ & $27 (22.9\%)$ & $2.103$ &  & $0.3494$ &  &  &  \\
Aware & $56 (83.6\%)$ & $52 (83.9\%)$ & $91 (77.1\%)$ &  &  &  &  &  &  \\
\multicolumn{10}{l}{\textit{Training on secure use of devices}} \\
Not aware & $8 (11.9\%)$ & $13 (21.0\%)$ & $29 (24.6\%)$ & $4.166$ &  & $0.1246$ &  &  &  \\
Aware & $58 (86.6\%)$ & $49 (79.0\%)$ & $88 (74.6\%)$ &  &  &  &  &  &  \\
\multicolumn{10}{l}{\textit{Multi-factor authentication (e.g., 2FA)}} \\
Not aware & $11 (16.4\%)$ & $19 (30.6\%)$ & $32 (27.1\%)$ & $3.740$ &  & $0.1541$ &  &  &  \\
Aware & $55 (82.1\%)$ & $43 (69.4\%)$ & $86 (72.9\%)$ &  &  &  &  &  &  \\
\multicolumn{10}{l}{\textit{Security operation center (SOC) 24/7}} \\
Not aware & $12 (17.9\%)$ & $18 (29.0\%)$ & $57 (48.3\%)$ & $18.626$ & $^{***}$ & $0.0000$ & $0.6023$ & $0.0001$ & $0.0269$ \\
Aware & $54 (80.6\%)$ & $44 (71.0\%)$ & $60 (50.8\%)$ &  &  &  &  &  &  \\
\multicolumn{10}{l}{\textit{Centralized management of software updates}} \\
Not aware & $11 (16.4\%)$ & $14 (22.6\%)$ & $33 (28.0\%)$ & $3.443$ &  & $0.1788$ &  &  &  \\
Aware & $56 (83.6\%)$ & $48 (77.4\%)$ & $83 (70.3\%)$ &  &  &  &  &  &  \\
\multicolumn{10}{l}{\textit{Organizational critical infrastructure access control}} \\
Not aware & $9 (13.4\%)$ & $12 (19.4\%)$ & $42 (35.6\%)$ & $12.949$ & $^{**}$ & $0.0015$ & $1.0000$ & $0.0025$ & $0.0437$ \\
Aware & $57 (85.1\%)$ & $50 (80.6\%)$ & $74 (62.7\%)$ &  &  &  &  &  &  \\
\bottomrule
\end{tabular}
\begin{flushleft}
We conducted Kruskal-Wallis tests to determine whether there are differences in awareness of solutions across groups. For solutions with significant differences across groups, we conducted post-hoc Dunn's tests with Bonferroni correction to determine which groups significantly differ from each other. Notes: $^{*} p < 0.05, ^{**} p < 0.01, ^{***} p < 0.001$.
\end{flushleft}
\end{table}

These results suggest that awareness of certain threats and solutions is positively associated with adoption of SOC albeit this association does not seem to be as diverse as its association with adoption of antimalware solutions.
Decision-makers in organizations adopting an internal SOC are more aware of one threat (i.e., online fraud) than decision-makers in organizations adopting an external SOC. There are no differences in their awareness of any solutions. Also, they are more aware of one threat (i.e., online fraud) than decision-makers in organizations which do not have a SOC.
Decision-makers in organizations adopting either realization of SOC (i.e., internal or external SOC) are more aware of two solutions (i.e., SOC and critical infrastructure access control) than decision-makers in organizations which do not have a SOC.

\subsection{RQ2a: Are there differences in cybersecurity awareness of decision-makers across groups based on organizational factors?}

\subsubsection{Organizational role type}

Table~\ref{table:role-t} presents the differences in respondents' \textit{awareness of threats} for groups based on their \textit{organizational role type}. The results indicate significant differences for awareness of six threats.
However, a single threat (i.e., industrial espionage) had clearly distinguishable differences between different organizational role types. Non-IT/IS executive decision-makers were significantly less aware of this threat than IT/IS executive and non-executive decision-makers.

\begin{table}[!htb]
\footnotesize 
\caption{\label{table:role-t} Differences in awareness of threats across organizational role types.}
\begin{tabular}{lrrrrlrrrr}
\toprule
 & 1: IT/IS executive & 2: Non-IT/IS executive & 3: Non-executive & $H$ &  & $p$ & $p_{1-2}$ & $p_{1-3}$ & $p_{2-3}$ \\ \midrule
\multicolumn{10}{l}{\textit{Loss of access to data (e.g., ransomware, locking of devices)}} \\
Not aware & $18 (17.1\%)$ & $30 (26.3\%)$ & $7 (10.9\%)$ & $6.912$ & $^{*}$ & $0.0316$ & $0.2413$ & $0.9727$ & $0.0358$ \\
Aware & $87 (82.9\%)$ & $83 (72.8\%)$ & $57 (89.1\%)$ &  &  &  &  &  &  \\
\multicolumn{10}{l}{\textit{Information system intrusion (e.g., hacking)}} \\
Not aware & $15 (14.3\%)$ & $24 (21.1\%)$ & $2 (3.1\%)$ & $10.602$ & $^{**}$ & $0.0050$ & $0.4678$ & $0.1378$ & $0.0034$ \\
Aware & $90 (85.7\%)$ & $90 (78.9\%)$ & $62 (96.9\%)$ &  &  &  &  &  &  \\
\multicolumn{10}{l}{\textit{Theft of business-critical data (industrial espionage)}} \\
Not aware & $22 (21.0\%)$ & $46 (40.4\%)$ & $9 (14.1\%)$ & $17.539$ & $^{***}$ & $0.0002$ & $0.0039$ & $0.9894$ & $0.0005$ \\
Aware & $83 (79.0\%)$ & $68 (59.6\%)$ & $55 (85.9\%)$ &  &  &  &  &  &  \\
\multicolumn{10}{l}{\textit{Distributed denial of service (DDoS) attacks}} \\
Not aware & $26 (24.8\%)$ & $50 (43.9\%)$ & $23 (35.9\%)$ & $9.064$ & $^{*}$ & $0.0108$ & $0.0079$ & $0.4216$ & $0.7998$ \\
Aware & $79 (75.2\%)$ & $63 (55.3\%)$ & $41 (64.1\%)$ &  &  &  &  &  &  \\
\multicolumn{10}{l}{\textit{Takeover of devices (e.g., botnets)}} \\
Not aware & $26 (24.8\%)$ & $44 (38.6\%)$ & $21 (32.8\%)$ & $4.795$ &  & $0.0910$ &  &  &  \\
Aware & $78 (74.3\%)$ & $69 (60.5\%)$ & $43 (67.2\%)$ &  &  &  &  &  &  \\
\multicolumn{10}{l}{\textit{Malware infection (e.g., viruses, worms, Trojans, spyware)}} \\
Not aware & $16 (15.2\%)$ & $20 (17.5\%)$ & $2 (3.1\%)$ & $7.691$ & $^{*}$ & $0.0214$ & $1.0000$ & $0.0814$ & $0.0210$ \\
Aware & $89 (84.8\%)$ & $93 (81.6\%)$ & $61 (95.3\%)$ &  &  &  &  &  &  \\
\multicolumn{10}{l}{\textit{Phishing}} \\
Not aware & $25 (23.8\%)$ & $37 (32.5\%)$ & $7 (10.9\%)$ & $10.288$ & $^{**}$ & $0.0058$ & $0.4117$ & $0.1775$ & $0.0041$ \\
Aware & $80 (76.2\%)$ & $77 (67.5\%)$ & $57 (89.1\%)$ &  &  &  &  &  &  \\
\multicolumn{10}{l}{\textit{Online fraud (e.g., CEO fraud, business email compromise)}} \\
Not aware & $25 (23.8\%)$ & $26 (22.8\%)$ & $8 (12.5\%)$ & $3.514$ &  & $0.1726$ &  &  &  \\
Aware & $80 (76.2\%)$ & $88 (77.2\%)$ & $56 (87.5\%)$ &  &  &  &  &  &  \\
\multicolumn{10}{l}{\textit{Internal threats (e.g., deliberate data deletion, unauthorized data access, unauthorized personal devices)}} \\
Not aware & $18 (17.1\%)$ & $27 (23.7\%)$ & $9 (14.1\%)$ & $3.503$ &  & $0.1735$ &  &  &  \\
Aware & $87 (82.9\%)$ & $82 (71.9\%)$ & $55 (85.9\%)$ &  &  &  &  &  &  \\
\bottomrule
\end{tabular}
\begin{flushleft}
We conducted Kruskal-Wallis tests to determine whether there are differences in awareness of threats across groups. For threats with significant differences across groups, we conducted post-hoc Dunn's tests with Bonferroni correction to determine which groups significantly differ from each other. Notes: $^{*} p < 0.05, ^{**} p < 0.01, ^{***} p < 0.001$.
\end{flushleft}
\end{table}

Table~\ref{table:role-s} presents the differences in respondents' \textit{awareness of solutions} for groups based on their \textit{organizational role type}. The results suggest significant differences for awareness of all 12 solutions. Nevertheless, only six had clear differences between different organizational role types.
For four solutions (i.e., training, multi-factor authentication, centralized management of software updates and critical infrastructure access control), non-IT/IS executive decision-makers were significantly less aware of these solutions than IT/IS executive and non-executive decision-makers.
Perhaps a bit surprisingly, non-executive decision-makers were significantly more aware of remote data deletion and cloud synchronization of data than both IT/IS and non-IT/IS executive decision-makers.

\begin{table}[!htb]
\footnotesize 
\caption{\label{table:role-s} Differences in awareness of solutions across organizational role types.}
\begin{tabular}{lrrrrlrrrr}
\toprule
 & 1: IT/IS executive & 2: Non-IT/IS executive & 3: Non-executive & $H$ &  & $p$ & $p_{1-2}$ & $p_{1-3}$ & $p_{2-3}$ \\ \midrule
\multicolumn{10}{l}{\textit{Remote data deletion on lost or stolen devices}} \\
Not aware & $16 (15.2\%)$ & $41 (36.0\%)$ & $13 (20.3\%)$ & $13.876$ & $^{***}$ & $0.0010$ & $0.0010$ & $1.0000$ & $0.0496$ \\
Aware & $88 (83.8\%)$ & $71 (62.3\%)$ & $51 (79.7\%)$ &  &  &  &  &  &  \\
\multicolumn{10}{l}{\textit{Advanced antimalware solutions (with EDR/XDR capabilities)}} \\
Not aware & $33 (31.4\%)$ & $46 (40.4\%)$ & $22 (34.4\%)$ & $1.901$ &  & $0.3866$ &  &  &  \\
Aware & $71 (67.6\%)$ & $67 (58.8\%)$ & $40 (62.5\%)$ &  &  &  &  &  &  \\
\multicolumn{10}{l}{\textit{Secure connection (e.g., VPN)}} \\
Not aware & $23 (21.9\%)$ & $26 (22.8\%)$ & $2 (3.1\%)$ & $12.408$ & $^{**}$ & $0.0020$ & $1.0000$ & $0.0063$ & $0.0032$ \\
Aware & $82 (78.1\%)$ & $88 (77.2\%)$ & $62 (96.9\%)$ &  &  &  &  &  &  \\
\multicolumn{10}{l}{\textit{Cloud synchronization of data}} \\
Not aware & $21 (20.0\%)$ & $25 (21.9\%)$ & $3 (4.7\%)$ & $9.365$ & $^{**}$ & $0.0093$ & $1.0000$ & $0.0304$ & $0.0109$ \\
Aware & $83 (79.0\%)$ & $89 (78.1\%)$ & $61 (95.3\%)$ &  &  &  &  &  &  \\
\multicolumn{10}{l}{\textit{Data backup}} \\
Not aware & $21 (20.0\%)$ & $23 (20.2\%)$ & $4 (6.3\%)$ & $6.586$ & $^{*}$ & $0.0372$ & $1.0000$ & $0.0693$ & $0.0545$ \\
Aware & $84 (80.0\%)$ & $90 (78.9\%)$ & $59 (92.2\%)$ &  &  &  &  &  &  \\
\multicolumn{10}{l}{\textit{Centralized device management, including mobile device management (MDM)}} \\
Not aware & $19 (18.1\%)$ & $37 (32.5\%)$ & $19 (29.7\%)$ & $6.157$ & $^{*}$ & $0.0460$ & $0.0487$ & $0.3145$ & $1.0000$ \\
Aware & $85 (81.0\%)$ & $76 (66.7\%)$ & $45 (70.3\%)$ &  &  &  &  &  &  \\
\multicolumn{10}{l}{\textit{Advanced firewalls (with IPS/IDS capabilities)}} \\
Not aware & $17 (16.2\%)$ & $30 (26.3\%)$ & $12 (18.8\%)$ & $3.491$ &  & $0.1746$ &  &  &  \\
Aware & $87 (82.9\%)$ & $84 (73.7\%)$ & $52 (81.3\%)$ &  &  &  &  &  &  \\
\multicolumn{10}{l}{\textit{Training on secure use of devices}} \\
Not aware & $22 (21.0\%)$ & $31 (27.2\%)$ & $7 (10.9\%)$ & $6.231$ & $^{*}$ & $0.0444$ & $0.8339$ & $0.3764$ & $0.0377$ \\
Aware & $82 (78.1\%)$ & $83 (72.8\%)$ & $56 (87.5\%)$ &  &  &  &  &  &  \\
\multicolumn{10}{l}{\textit{Multi-factor authentication (e.g., 2FA)}} \\
Not aware & $21 (20.0\%)$ & $40 (35.1\%)$ & $14 (21.9\%)$ & $7.101$ & $^{*}$ & $0.0287$ & $0.0392$ & $1.0000$ & $0.1680$ \\
Aware & $83 (79.0\%)$ & $74 (64.9\%)$ & $50 (78.1\%)$ &  &  &  &  &  &  \\
\multicolumn{10}{l}{\textit{Security operation center (SOC) 24/7}} \\
Not aware & $27 (25.7\%)$ & $54 (47.4\%)$ & $27 (42.2\%)$ & $11.836$ & $^{**}$ & $0.0027$ & $0.0025$ & $0.0818$ & $1.0000$ \\
Aware & $78 (74.3\%)$ & $59 (51.8\%)$ & $36 (56.3\%)$ &  &  &  &  &  &  \\
\multicolumn{10}{l}{\textit{Centralized management of software updates}} \\
Not aware & $23 (21.9\%)$ & $41 (36.0\%)$ & $10 (15.6\%)$ & $10.899$ & $^{**}$ & $0.0043$ & $0.0425$ & $1.0000$ & $0.0072$ \\
Aware & $82 (78.1\%)$ & $71 (62.3\%)$ & $54 (84.4\%)$ &  &  &  &  &  &  \\
\multicolumn{10}{l}{\textit{Organizational critical infrastructure access control}} \\
Not aware & $19 (18.1\%)$ & $47 (41.2\%)$ & $13 (20.3\%)$ & $17.487$ & $^{***}$ & $0.0002$ & $0.0003$ & $1.0000$ & $0.0080$ \\
Aware & $86 (81.9\%)$ & $65 (57.0\%)$ & $50 (78.1\%)$ &  &  &  &  &  &  \\
\bottomrule
\end{tabular}
\begin{flushleft}
We conducted Kruskal-Wallis tests to determine whether there are differences in awareness of solutions across groups. For solutions with significant differences across groups, we conducted post-hoc Dunn's tests with Bonferroni correction to determine which groups significantly differ from each other. Notes: $^{*} p < 0.05, ^{**} p < 0.01, ^{***} p < 0.001$.
\end{flushleft}
\end{table}

These results indicate some differences in decision-makers' awareness of certain threats and solutions depending on their organizational role type.
Non-IT/IS executive decision-makers are less aware of one threat (i.e., industrial espionage) and four solutions (i.e., training, multi-factor authentication, centralized management of software updates and critical infrastructure access control) than both IT/IS executive and non-executive decision-makers. They are also less aware of two solutions (i.e., remote data deletion and cloud synchronization of data) than non-executive decision-makers.
IT/IS executive decision-makers are additionally less aware of these two solutions than non-executive decision-makers.

\subsubsection{Organization size}

Table~\ref{table:size-t} shows the differences in respondents' \textit{awareness of threats} for groups based on \textit{organization size}. The results indicate significant differences for awareness of a single threat. Post-hoc tests did not reveal any organization sizes that would be clearly distinguishable from others.
    
\begin{table}[!htb]
\footnotesize 
\caption{\label{table:size-t} Differences in awareness of threats across organization size groups.}
\begin{tabular}{lrrrrrlrrrrrrr}
\toprule
 & 1: Micro & 2: Small & 3: Medium & 4: Large & $H$ &  & $p$ & $p_{1-2}$ & $p_{1-3}$ & $p_{2-3}$ & $p_{1-4}$ & $p_{2-4}$ & $p_{3-4}$ \\ \midrule
\multicolumn{14}{l}{\textit{Loss of access to data (e.g., ransomware, locking of devices)}} \\
Not aware & $27 (20.3\%)$ & $20 (22.7\%)$ & $6 (12.2\%)$ & $2 (15.4\%)$ & $2.503$ &  & $0.4748$ &  &  &  &  &  &  \\
Aware & $106 (79.7\%)$ & $67 (76.1\%)$ & $43 (87.8\%)$ & $11 (84.6\%)$ &  &  &  &  &  &  &  &  &  \\
\multicolumn{14}{l}{\textit{Information system intrusion (e.g., hacking)}} \\
Not aware & $22 (16.5\%)$ & $12 (13.6\%)$ & $6 (12.2\%)$ & $1 (7.7\%)$ & $1.184$ &  & $0.7569$ &  &  &  &  &  &  \\
Aware & $111 (83.5\%)$ & $76 (86.4\%)$ & $43 (87.8\%)$ & $12 (92.3\%)$ &  &  &  &  &  &  &  &  &  \\
\multicolumn{14}{l}{\textit{Theft of business-critical data (industrial espionage)}} \\
Not aware & $42 (31.6\%)$ & $26 (29.5\%)$ & $8 (16.3\%)$ & $1 (7.7\%)$ & $6.931$ &  & $0.0742$ &  &  &  &  &  &  \\
Aware & $91 (68.4\%)$ & $62 (70.5\%)$ & $41 (83.7\%)$ & $12 (92.3\%)$ &  &  &  &  &  &  &  &  &  \\
\multicolumn{14}{l}{\textit{Distributed denial of service (DDoS) attacks}} \\
Not aware & $55 (41.4\%)$ & $26 (29.5\%)$ & $13 (26.5\%)$ & $5 (38.5\%)$ & $4.948$ &  & $0.1757$ &  &  &  &  &  &  \\
Aware & $78 (58.6\%)$ & $61 (69.3\%)$ & $36 (73.5\%)$ & $8 (61.5\%)$ &  &  &  &  &  &  &  &  &  \\
\multicolumn{14}{l}{\textit{Takeover of devices (e.g., botnets)}} \\
Not aware & $56 (42.1\%)$ & $17 (19.3\%)$ & $12 (24.5\%)$ & $6 (46.2\%)$ & $14.460$ & $^{**}$ & $0.0023$ & $0.0034$ & $0.1472$ & $1.0000$ & $1.0000$ & $0.3513$ & $0.8311$ \\
Aware & $77 (57.9\%)$ & $69 (78.4\%)$ & $37 (75.5\%)$ & $7 (53.8\%)$ &  &  &  &  &  &  &  &  &  \\
\multicolumn{14}{l}{\textit{Malware infection (e.g., viruses, worms, Trojans, spyware)}} \\
Not aware & $19 (14.3\%)$ & $13 (14.8\%)$ & $6 (12.2\%)$ & $ (0.0\%)$ & $2.284$ &  & $0.5156$ &  &  &  &  &  &  \\
Aware & $114 (85.7\%)$ & $74 (84.1\%)$ & $42 (85.7\%)$ & $13 (100.0\%)$ &  &  &  &  &  &  &  &  &  \\
\multicolumn{14}{l}{\textit{Phishing}} \\
Not aware & $39 (29.3\%)$ & $19 (21.6\%)$ & $10 (20.4\%)$ & $1 (7.7\%)$ & $4.501$ &  & $0.2122$ &  &  &  &  &  &  \\
Aware & $94 (70.7\%)$ & $69 (78.4\%)$ & $39 (79.6\%)$ & $12 (92.3\%)$ &  &  &  &  &  &  &  &  &  \\
\multicolumn{14}{l}{\textit{Online fraud (e.g., CEO fraud, business email compromise)}} \\
Not aware & $32 (24.1\%)$ & $15 (17.0\%)$ & $9 (18.4\%)$ & $3 (23.1\%)$ & $1.818$ &  & $0.6110$ &  &  &  &  &  &  \\
Aware & $101 (75.9\%)$ & $73 (83.0\%)$ & $40 (81.6\%)$ & $10 (76.9\%)$ &  &  &  &  &  &  &  &  &  \\
\multicolumn{14}{l}{\textit{Internal threats (e.g., deliberate data deletion, unauthorized data access, unauthorized personal devices)}} \\
Not aware & $33 (24.8\%)$ & $14 (15.9\%)$ & $5 (10.2\%)$ & $2 (15.4\%)$ & $6.515$ &  & $0.0891$ &  &  &  &  &  &  \\
Aware & $96 (72.2\%)$ & $73 (83.0\%)$ & $44 (89.8\%)$ & $11 (84.6\%)$ &  &  &  &  &  &  &  &  &  \\
\bottomrule
\end{tabular}
\begin{flushleft}
We conducted Kruskal-Wallis tests to determine whether there are differences in awareness of threats across groups. For threats with significant differences across groups, we conducted post-hoc Dunn's tests with Bonferroni correction to determine which groups significantly differ from each other. Notes: $^{*} p < 0.05, ^{**} p < 0.01, ^{***} p < 0.001$.
\end{flushleft}
\end{table}

Table~\ref{table:size-s} presents the differences in respondents' \textit{awareness of solutions} for groups based on \textit{organization size}. The results indicate significant differences for awareness of four solutions. Again, no organization size was clearly distinguishable from others for any of these solutions.

\begin{table}[!htb]
\footnotesize 
\caption{\label{table:size-s} Differences in awareness of solutions across organization size groups.}
\begin{tabular}{lrrrrrlrrrrrrr}
\toprule
 & 1: Micro & 2: Small & 3: Medium & 4: Large & $H$ &  & $p$ & $p_{1-2}$ & $p_{1-3}$ & $p_{2-3}$ & $p_{1-4}$ & $p_{2-4}$ & $p_{3-4}$ \\ \midrule
\multicolumn{14}{l}{\textit{Remote data deletion on lost or stolen devices}} \\
Not aware & $43 (32.3\%)$ & $22 (25.0\%)$ & $2 (4.1\%)$ & $3 (23.1\%)$ & $15.127$ & $^{**}$ & $0.0017$ & $1.0000$ & $0.0006$ & $0.0406$ & $1.0000$ & $1.0000$ & $0.9794$ \\
Aware & $89 (66.9\%)$ & $65 (73.9\%)$ & $46 (93.9\%)$ & $10 (76.9\%)$ &  &  &  &  &  &  &  &  &  \\
\multicolumn{14}{l}{\textit{Advanced antimalware solutions (with EDR/XDR capabilities)}} \\
Not aware & $57 (42.9\%)$ & $29 (33.0\%)$ & $8 (16.3\%)$ & $7 (53.8\%)$ & $12.977$ & $^{**}$ & $0.0047$ & $0.7582$ & $0.0057$ & $0.3251$ & $1.0000$ & $0.9113$ & $0.0811$ \\
Aware & $74 (55.6\%)$ & $58 (65.9\%)$ & $40 (81.6\%)$ & $6 (46.2\%)$ &  &  &  &  &  &  &  &  &  \\
\multicolumn{14}{l}{\textit{Secure connection (e.g., VPN)}} \\
Not aware & $26 (19.5\%)$ & $19 (21.6\%)$ & $6 (12.2\%)$ & $ (0.0\%)$ & $4.916$ &  & $0.1780$ &  &  &  &  &  &  \\
Aware & $107 (80.5\%)$ & $69 (78.4\%)$ & $43 (87.8\%)$ & $13 (100.0\%)$ &  &  &  &  &  &  &  &  &  \\
\multicolumn{14}{l}{\textit{Cloud synchronization of data}} \\
Not aware & $27 (20.3\%)$ & $14 (15.9\%)$ & $7 (14.3\%)$ & $1 (7.7\%)$ & $2.171$ &  & $0.5378$ &  &  &  &  &  &  \\
Aware & $105 (78.9\%)$ & $74 (84.1\%)$ & $42 (85.7\%)$ & $12 (92.3\%)$ &  &  &  &  &  &  &  &  &  \\
\multicolumn{14}{l}{\textit{Data backup}} \\
Not aware & $19 (14.3\%)$ & $18 (20.5\%)$ & $11 (22.4\%)$ & $ (0.0\%)$ & $5.193$ &  & $0.1582$ &  &  &  &  &  &  \\
Aware & $113 (85.0\%)$ & $70 (79.5\%)$ & $37 (75.5\%)$ & $13 (100.0\%)$ &  &  &  &  &  &  &  &  &  \\
\multicolumn{14}{l}{\textit{Centralized device management, including mobile device management (MDM)}} \\
Not aware & $49 (36.8\%)$ & $15 (17.0\%)$ & $7 (14.3\%)$ & $4 (30.8\%)$ & $15.219$ & $^{**}$ & $0.0016$ & $0.0070$ & $0.0124$ & $1.0000$ & $1.0000$ & $1.0000$ & $1.0000$ \\
Aware & $83 (62.4\%)$ & $72 (81.8\%)$ & $42 (85.7\%)$ & $9 (69.2\%)$ &  &  &  &  &  &  &  &  &  \\
\multicolumn{14}{l}{\textit{Advanced firewalls (with IPS/IDS capabilities)}} \\
Not aware & $34 (25.6\%)$ & $15 (17.0\%)$ & $7 (14.3\%)$ & $3 (23.1\%)$ & $3.992$ &  & $0.2624$ &  &  &  &  &  &  \\
Aware & $98 (73.7\%)$ & $73 (83.0\%)$ & $42 (85.7\%)$ & $10 (76.9\%)$ &  &  &  &  &  &  &  &  &  \\
\multicolumn{14}{l}{\textit{Training on secure use of devices}} \\
Not aware & $33 (24.8\%)$ & $20 (22.7\%)$ & $7 (14.3\%)$ & $ (0.0\%)$ & $5.963$ &  & $0.1134$ &  &  &  &  &  &  \\
Aware & $99 (74.4\%)$ & $68 (77.3\%)$ & $41 (83.7\%)$ & $13 (100.0\%)$ &  &  &  &  &  &  &  &  &  \\
\multicolumn{14}{l}{\textit{Multi-factor authentication (e.g., 2FA)}} \\
Not aware & $43 (32.3\%)$ & $18 (20.5\%)$ & $12 (24.5\%)$ & $2 (15.4\%)$ & $5.048$ &  & $0.1683$ &  &  &  &  &  &  \\
Aware & $89 (66.9\%)$ & $70 (79.5\%)$ & $37 (75.5\%)$ & $11 (84.6\%)$ &  &  &  &  &  &  &  &  &  \\
\multicolumn{14}{l}{\textit{Security operation center (SOC) 24/7}} \\
Not aware & $62 (46.6\%)$ & $28 (31.8\%)$ & $11 (22.4\%)$ & $7 (53.8\%)$ & $11.349$ & $^{**}$ & $0.0100$ & $0.1903$ & $0.0232$ & $1.0000$ & $1.0000$ & $0.8095$ & $0.2541$ \\
Aware & $71 (53.4\%)$ & $59 (67.0\%)$ & $37 (75.5\%)$ & $6 (46.2\%)$ &  &  &  &  &  &  &  &  &  \\
\multicolumn{14}{l}{\textit{Centralized management of software updates}} \\
Not aware & $44 (33.1\%)$ & $20 (22.7\%)$ & $8 (16.3\%)$ & $2 (15.4\%)$ & $7.143$ &  & $0.0675$ &  &  &  &  &  &  \\
Aware & $88 (66.2\%)$ & $67 (76.1\%)$ & $41 (83.7\%)$ & $11 (84.6\%)$ &  &  &  &  &  &  &  &  &  \\
\multicolumn{14}{l}{\textit{Organizational critical infrastructure access control}} \\
Not aware & $45 (33.8\%)$ & $18 (20.5\%)$ & $11 (22.4\%)$ & $5 (38.5\%)$ & $5.646$ &  & $0.1302$ &  &  &  &  &  &  \\
Aware & $88 (66.2\%)$ & $68 (77.3\%)$ & $37 (75.5\%)$ & $8 (61.5\%)$ &  &  &  &  &  &  &  &  &  \\
\bottomrule
\end{tabular}
\begin{flushleft}
We conducted Kruskal-Wallis tests to determine whether there are differences in awareness of solutions across groups. For solutions with significant differences across groups, we conducted post-hoc Dunn's tests with Bonferroni correction to determine which groups significantly differ from each other. Notes: $^{*} p < 0.05, ^{**} p < 0.01, ^{***} p < 0.001$.
\end{flushleft}
\end{table}

Based on the above, we can conclude that organization size is not associated with awareness of neither threats nor solutions.

\subsection{RQ2b: Are there differences in cybersecurity awareness of decision-makers across groups based on their personal characteristics?}

\subsubsection{Gender}

Table~\ref{table:sex-t} shows the differences in respondents' \textit{awareness of threats} for groups based on their \textit{gender}. The results indicate significant differences in awareness of five threats. Male respondents were significantly more aware of loss of access to data, industrial espionage, DDoS attacks, botnets and phishing than female respondents.

\begin{table}[!htb]
\footnotesize 
\caption{\label{table:sex-t} Differences in awareness of threats across genders.}
\begin{tabular}{lrrrlr}
\toprule
 & Female & Male & $W$ &  & $p$ \\ \midrule
\multicolumn{6}{l}{\textit{Loss of access to data (e.g., ransomware, locking of devices)}} \\
Not aware & $25 (28.7\%)$ & $30 (15.4\%)$ & $7319.0$ & $^{**}$ & $0.0097$ \\
Aware & $62 (71.3\%)$ & $164 (84.1\%)$ &  &  &  \\
\multicolumn{6}{l}{\textit{Information system intrusion (e.g., hacking)}} \\
Not aware & $10 (11.5\%)$ & $31 (15.9\%)$ & $8856.0$ &  & $0.3341$ \\
Aware & $77 (88.5\%)$ & $164 (84.1\%)$ &  &  &  \\
\multicolumn{6}{l}{\textit{Theft of business-critical data (industrial espionage)}} \\
Not aware & $31 (35.6\%)$ & $46 (23.6\%)$ & $7461.0$ & $^{*}$ & $0.0365$ \\
Aware & $56 (64.4\%)$ & $149 (76.4\%)$ &  &  &  \\
\multicolumn{6}{l}{\textit{Distributed denial of service (DDoS) attacks}} \\
Not aware & $43 (49.4\%)$ & $56 (28.7\%)$ & $6704.0$ & $^{***}$ & $0.0009$ \\
Aware & $44 (50.6\%)$ & $138 (70.8\%)$ &  &  &  \\
\multicolumn{6}{l}{\textit{Takeover of devices (e.g., botnets)}} \\
Not aware & $41 (47.1\%)$ & $50 (25.6\%)$ & $6614.0$ & $^{***}$ & $0.0005$ \\
Aware & $46 (52.9\%)$ & $143 (73.3\%)$ &  &  &  \\
\multicolumn{6}{l}{\textit{Malware infection (e.g., viruses, worms, Trojans, spyware)}} \\
Not aware & $16 (18.4\%)$ & $22 (11.3\%)$ & $7808.5$ &  & $0.1149$ \\
Aware & $71 (81.6\%)$ & $171 (87.7\%)$ &  &  &  \\
\multicolumn{6}{l}{\textit{Phishing}} \\
Not aware & $32 (36.8\%)$ & $37 (19.0\%)$ & $6972.0$ & $^{**}$ & $0.0013$ \\
Aware & $55 (63.2\%)$ & $158 (81.0\%)$ &  &  &  \\
\multicolumn{6}{l}{\textit{Online fraud (e.g., CEO fraud, business email compromise)}} \\
Not aware & $19 (21.8\%)$ & $40 (20.5\%)$ & $8370.0$ &  & $0.8016$ \\
Aware & $68 (78.2\%)$ & $155 (79.5\%)$ &  &  &  \\
\multicolumn{6}{l}{\textit{Internal threats (e.g., deliberate data deletion, unauthorized data access, unauthorized personal devices)}} \\
Not aware & $18 (20.7\%)$ & $36 (18.5\%)$ & $8042.0$ &  & $0.6871$ \\
Aware & $68 (78.2\%)$ & $155 (79.5\%)$ &  &  &  \\
\bottomrule
\end{tabular}
\begin{flushleft}
We conducted Wilcoxon tests to determine whether there are differences in awareness of threats across groups. Notes: $^{*} p < 0.05, ^{**} p < 0.01, ^{***} p < 0.001$.
\end{flushleft}
\end{table}

Table~\ref{table:sex-s} shows the differences in respondents' \textit{awareness of solutions} for groups based on their \textit{gender}. The results indicate significant differences for awareness of six solutions. Male respondents were significantly more aware of advanced antimalware solutions with EDR/XDR capabilities, centralized device management, training, multi-factor authentication, centralized management of software updates and critical infrastructure access control than female respondents.

\begin{table}[!htb]
\footnotesize 
\caption{\label{table:sex-s} Differences in awareness of solutions across genders.}
\begin{tabular}{lrrrlr}
\toprule
 & Female & Male & $W$ &  & $p$ \\ \midrule
\multicolumn{6}{l}{\textit{Remote data deletion on lost or stolen devices}} \\
Not aware & $26 (29.9\%)$ & $44 (22.6\%)$ & $7682.0$ &  & $0.1871$ \\
Aware & $60 (69.0\%)$ & $149 (76.4\%)$ &  &  &  \\
\multicolumn{6}{l}{\textit{Advanced antimalware solutions (with EDR/XDR capabilities)}} \\
Not aware & $41 (47.1\%)$ & $60 (30.8\%)$ & $6900.0$ & $^{**}$ & $0.0086$ \\
Aware & $45 (51.7\%)$ & $132 (67.7\%)$ &  &  &  \\
\multicolumn{6}{l}{\textit{Secure connection (e.g., VPN)}} \\
Not aware & $18 (20.7\%)$ & $33 (16.9\%)$ & $8163.0$ &  & $0.4494$ \\
Aware & $69 (79.3\%)$ & $162 (83.1\%)$ &  &  &  \\
\multicolumn{6}{l}{\textit{Cloud synchronization of data}} \\
Not aware & $16 (18.4\%)$ & $33 (16.9\%)$ & $8322.5$ &  & $0.7793$ \\
Aware & $71 (81.6\%)$ & $161 (82.6\%)$ &  &  &  \\
\multicolumn{6}{l}{\textit{Data backup}} \\
Not aware & $16 (18.4\%)$ & $32 (16.4\%)$ & $8166.0$ &  & $0.6671$ \\
Aware & $70 (80.5\%)$ & $162 (83.1\%)$ &  &  &  \\
\multicolumn{6}{l}{\textit{Centralized device management, including mobile device management (MDM)}} \\
Not aware & $32 (36.8\%)$ & $43 (22.1\%)$ & $7087.0$ & $^{**}$ & $0.0089$ \\
Aware & $54 (62.1\%)$ & $151 (77.4\%)$ &  &  &  \\
\multicolumn{6}{l}{\textit{Advanced firewalls (with IPS/IDS capabilities)}} \\
Not aware & $21 (24.1\%)$ & $38 (19.5\%)$ & $8055.0$ &  & $0.3880$ \\
Aware & $66 (75.9\%)$ & $156 (80.0\%)$ &  &  &  \\
\multicolumn{6}{l}{\textit{Training on secure use of devices}} \\
Not aware & $27 (31.0\%)$ & $33 (16.9\%)$ & $7225.5$ & $^{**}$ & $0.0087$ \\
Aware & $60 (69.0\%)$ & $160 (82.1\%)$ &  &  &  \\
\multicolumn{6}{l}{\textit{Multi-factor authentication (e.g., 2FA)}} \\
Not aware & $33 (37.9\%)$ & $42 (21.5\%)$ & $7065.0$ & $^{**}$ & $0.0044$ \\
Aware & $54 (62.1\%)$ & $152 (77.9\%)$ &  &  &  \\
\multicolumn{6}{l}{\textit{Security operation center (SOC) 24/7}} \\
Not aware & $39 (44.8\%)$ & $69 (35.4\%)$ & $7633.5$ &  & $0.1498$ \\
Aware & $48 (55.2\%)$ & $124 (63.6\%)$ &  &  &  \\
\multicolumn{6}{l}{\textit{Centralized management of software updates}} \\
Not aware & $34 (39.1\%)$ & $40 (20.5\%)$ & $6764.0$ & $^{***}$ & $0.0010$ \\
Aware & $52 (59.8\%)$ & $154 (79.0\%)$ &  &  &  \\
\multicolumn{6}{l}{\textit{Organizational critical infrastructure access control}} \\
Not aware & $34 (39.1\%)$ & $45 (23.1\%)$ & $6953.0$ & $^{**}$ & $0.0056$ \\
Aware & $52 (59.8\%)$ & $148 (75.9\%)$ &  &  &  \\
\bottomrule
\end{tabular}
\begin{flushleft}
We conducted Wilcoxon tests to determine whether there are differences in awareness of threats across groups. Notes: $^{*} p < 0.05, ^{**} p < 0.01, ^{***} p < 0.001$.
\end{flushleft}
\end{table}

These results suggest that male decision-makers are more aware of certain threats and solutions than their female counterparts. To be more specific, male decision-makers were more aware of five out of nine threats ($55.6\%$), and six out of 12 solutions ($50.0\%$).

\subsubsection{Formal education}

Table~\ref{table:edu-t} shows the differences in respondents' \textit{awareness of threats} for groups based on \textit{formal education}. The results do not suggest any significant differences in awareness of threats among these groups.

\begin{table}[!htb]
\footnotesize 
\caption{\label{table:edu-t} Differences in awareness of threats across formal education groups.}
\begin{tabular}{lrrrrrr}
\toprule
 & Finished high school & Bachelor’s degree & Master’s degree & PhD degree & $H$ & $p$ \\ \midrule
\multicolumn{7}{l}{\textit{Loss of access to data (e.g., ransomware, locking of devices)}} \\
Not aware & $11 (20.4\%)$ & $15 (16.7\%)$ & $22 (20.8\%)$ & $7 (21.2\%)$ & $0.687$ & $0.8764$ \\
Aware & $43 (79.6\%)$ & $75 (83.3\%)$ & $83 (78.3\%)$ & $26 (78.8\%)$ &  &  \\
\multicolumn{7}{l}{\textit{Information system intrusion (e.g., hacking)}} \\
Not aware & $6 (11.1\%)$ & $9 (10.0\%)$ & $20 (18.9\%)$ & $6 (18.2\%)$ & $3.951$ & $0.2668$ \\
Aware & $48 (88.9\%)$ & $81 (90.0\%)$ & $86 (81.1\%)$ & $27 (81.8\%)$ &  &  \\
\multicolumn{7}{l}{\textit{Theft of business-critical data (industrial espionage)}} \\
Not aware & $15 (27.8\%)$ & $24 (26.7\%)$ & $24 (22.6\%)$ & $14 (42.4\%)$ & $4.978$ & $0.1734$ \\
Aware & $39 (72.2\%)$ & $66 (73.3\%)$ & $82 (77.4\%)$ & $19 (57.6\%)$ &  &  \\
\multicolumn{7}{l}{\textit{Distributed denial of service (DDoS) attacks}} \\
Not aware & $21 (38.9\%)$ & $33 (36.7\%)$ & $36 (34.0\%)$ & $9 (27.3\%)$ & $1.350$ & $0.7172$ \\
Aware & $33 (61.1\%)$ & $57 (63.3\%)$ & $69 (65.1\%)$ & $24 (72.7\%)$ &  &  \\
\multicolumn{7}{l}{\textit{Takeover of devices (e.g., botnets)}} \\
Not aware & $19 (35.2\%)$ & $29 (32.2\%)$ & $32 (30.2\%)$ & $11 (33.3\%)$ & $0.382$ & $0.9440$ \\
Aware & $35 (64.8\%)$ & $60 (66.7\%)$ & $73 (68.9\%)$ & $22 (66.7\%)$ &  &  \\
\multicolumn{7}{l}{\textit{Malware infection (e.g., viruses, worms, Trojans, spyware)}} \\
Not aware & $10 (18.5\%)$ & $10 (11.1\%)$ & $12 (11.3\%)$ & $6 (18.2\%)$ & $2.548$ & $0.4667$ \\
Aware & $44 (81.5\%)$ & $79 (87.8\%)$ & $93 (87.7\%)$ & $27 (81.8\%)$ &  &  \\
\multicolumn{7}{l}{\textit{Phishing}} \\
Not aware & $19 (35.2\%)$ & $18 (20.0\%)$ & $22 (20.8\%)$ & $10 (30.3\%)$ & $5.719$ & $0.1261$ \\
Aware & $35 (64.8\%)$ & $72 (80.0\%)$ & $84 (79.2\%)$ & $23 (69.7\%)$ &  &  \\
\multicolumn{7}{l}{\textit{Online fraud (e.g., CEO fraud, business email compromise)}} \\
Not aware & $15 (27.8\%)$ & $17 (18.9\%)$ & $22 (20.8\%)$ & $5 (15.2\%)$ & $2.422$ & $0.4896$ \\
Aware & $39 (72.2\%)$ & $73 (81.1\%)$ & $84 (79.2\%)$ & $28 (84.8\%)$ &  &  \\
\multicolumn{7}{l}{\textit{Internal threats (e.g., deliberate data deletion, unauthorized data access, unauthorized personal devices)}} \\
Not aware & $13 (24.1\%)$ & $14 (15.6\%)$ & $21 (19.8\%)$ & $6 (18.2\%)$ & $1.807$ & $0.6135$ \\
Aware & $40 (74.1\%)$ & $76 (84.4\%)$ & $82 (77.4\%)$ & $26 (78.8\%)$ &  &  \\
\bottomrule
\end{tabular}
\begin{flushleft}
We conducted Kruskal-Wallis tests to determine whether there are differences in awareness of threats across groups.
\end{flushleft}
\end{table}

Table~\ref{table:edu-s} shows the differences in respondents' \textit{awareness of solutions} for groups based on \textit{formal education}. These results also do not suggest any significant differences in awareness of solutions among these groups.

\begin{table}[!htb]
\footnotesize 
\caption{\label{table:edu-s} Differences in awareness of solutions across formal education groups.}
\begin{tabular}{lrrrrrlr}
\toprule
 & Finished high school & Bachelor’s degree & Master’s degree & PhD degree & $H$ & $p$ \\ \midrule
\multicolumn{7}{l}{\textit{Remote data deletion on lost or stolen devices}} \\
Not aware & $17 (31.5\%)$ & $21 (23.3\%)$ & $24 (22.6\%)$ & $8 (24.2\%)$ & $1.532$ & $0.6749$ \\
Aware & $37 (68.5\%)$ & $67 (74.4\%)$ & $81 (76.4\%)$ & $25 (75.8\%)$ &  &  \\
\multicolumn{7}{l}{\textit{Advanced antimalware solutions (with EDR/XDR capabilities)}} \\
Not aware & $26 (48.1\%)$ & $32 (35.6\%)$ & $33 (31.1\%)$ & $10 (30.3\%)$ & $5.307$ & $0.1507$ \\
Aware & $27 (50.0\%)$ & $56 (62.2\%)$ & $72 (67.9\%)$ & $23 (69.7\%)$ &  &  \\
\multicolumn{7}{l}{\textit{Secure connection (e.g., VPN)}} \\
Not aware & $11 (20.4\%)$ & $17 (18.9\%)$ & $18 (17.0\%)$ & $5 (15.2\%)$ & $0.507$ & $0.9173$ \\
Aware & $43 (79.6\%)$ & $73 (81.1\%)$ & $88 (83.0\%)$ & $28 (84.8\%)$ &  &  \\
\multicolumn{7}{l}{\textit{Cloud synchronization of data}} \\
Not aware & $8 (14.8\%)$ & $17 (18.9\%)$ & $18 (17.0\%)$ & $6 (18.2\%)$ & $0.456$ & $0.9284$ \\
Aware & $46 (85.2\%)$ & $72 (80.0\%)$ & $88 (83.0\%)$ & $27 (81.8\%)$ &  &  \\
\multicolumn{7}{l}{\textit{Data backup}} \\
Not aware & $6 (11.1\%)$ & $11 (12.2\%)$ & $22 (20.8\%)$ & $9 (27.3\%)$ & $6.071$ & $0.1082$ \\
Aware & $48 (88.9\%)$ & $77 (85.6\%)$ & $84 (79.2\%)$ & $24 (72.7\%)$ &  &  \\
\multicolumn{7}{l}{\textit{Centralized device management, including mobile device management (MDM)}} \\
Not aware & $20 (37.0\%)$ & $21 (23.3\%)$ & $26 (24.5\%)$ & $8 (24.2\%)$ & $4.080$ & $0.2529$ \\
Aware & $33 (61.1\%)$ & $68 (75.6\%)$ & $80 (75.5\%)$ & $25 (75.8\%)$ &  &  \\
\multicolumn{7}{l}{\textit{Advanced firewalls (with IPS/IDS capabilities)}} \\
Not aware & $11 (20.4\%)$ & $17 (18.9\%)$ & $23 (21.7\%)$ & $8 (24.2\%)$ & $0.445$ & $0.9308$ \\
Aware & $43 (79.6\%)$ & $72 (80.0\%)$ & $83 (78.3\%)$ & $25 (75.8\%)$ &  &  \\
\multicolumn{7}{l}{\textit{Training on secure use of devices}} \\
Not aware & $15 (27.8\%)$ & $15 (16.7\%)$ & $20 (18.9\%)$ & $10 (30.3\%)$ & $4.248$ & $0.2359$ \\
Aware & $39 (72.2\%)$ & $73 (81.1\%)$ & $86 (81.1\%)$ & $23 (69.7\%)$ &  &  \\
\multicolumn{7}{l}{\textit{Multi-factor authentication (e.g., 2FA)}} \\
Not aware & $15 (27.8\%)$ & $28 (31.1\%)$ & $24 (22.6\%)$ & $8 (24.2\%)$ & $2.053$ & $0.5615$ \\
Aware & $39 (72.2\%)$ & $61 (67.8\%)$ & $82 (77.4\%)$ & $25 (75.8\%)$ &  &  \\
\multicolumn{7}{l}{\textit{Security operation center (SOC) 24/7}} \\
Not aware & $25 (46.3\%)$ & $32 (35.6\%)$ & $43 (40.6\%)$ & $8 (24.2\%)$ & $4.715$ & $0.1939$ \\
Aware & $29 (53.7\%)$ & $57 (63.3\%)$ & $62 (58.5\%)$ & $25 (75.8\%)$ &  &  \\
\multicolumn{7}{l}{\textit{Centralized management of software updates}} \\
Not aware & $17 (31.5\%)$ & $17 (18.9\%)$ & $32 (30.2\%)$ & $8 (24.2\%)$ & $4.172$ & $0.2435$ \\
Aware & $36 (66.7\%)$ & $72 (80.0\%)$ & $74 (69.8\%)$ & $25 (75.8\%)$ &  &  \\
\multicolumn{7}{l}{\textit{Organizational critical infrastructure access control}} \\
Not aware & $18 (33.3\%)$ & $21 (23.3\%)$ & $32 (30.2\%)$ & $8 (24.2\%)$ & $2.316$ & $0.5095$ \\
Aware & $35 (64.8\%)$ & $68 (75.6\%)$ & $73 (68.9\%)$ & $25 (75.8\%)$ &  &  \\
\bottomrule
\end{tabular}
\begin{flushleft}
We conducted Kruskal-Wallis tests to determine whether there are differences in awareness of solutions across groups.
\end{flushleft}
\end{table}

Based on the above, we can conclude that formal education is not associated with cybersecurity awareness of decision-makers.

\subsubsection{Age}

Table~\ref{table:age-t} shows the differences in respondents' \textit{age} for groups based on \textit{awareness of threats}. The results reveal significant differences in average age for awareness of three threats (i.e., industrial espionage, malware infection and phishing). Respondents who were aware of these threats were significantly older than those who were not.

\begin{table}[!htb]
\footnotesize 
\caption{\label{table:age-t} Differences in age across groups based on awareness of threats.}
\begin{tabular}{lrrrlr}
\toprule
 & Not aware & Aware & $t$ &  & $p$ \\ \midrule
Loss of access to data (e.g., ransomware, locking of devices) & $34.42$ & $37.24$ & $-1.557$ &  & $0.1205$ \\
Information system intrusion (e.g., hacking) & $35.66$ & $36.83$ & $-0.576$ &  & $0.5650$ \\
Theft of business-critical data (industrial espionage) & $33.55$ & $37.83$ & $-2.689$ & $^{**}$ & $0.0076$ \\
Distributed denial of service (DDoS) attacks & $37.31$ & $36.35$ & $0.640$ &  & $0.5224$ \\
Takeover of devices (e.g., botnets) & $35.54$ & $37.23$ & $-1.090$ &  & $0.2766$ \\
Malware infection (e.g., viruses, worms, Trojans, spyware) & $32.37$ & $37.26$ & $-2.354$ & $^{*}$ & $0.0193$ \\
Phishing & $33.81$ & $37.59$ & $-2.276$ & $^{*}$ & $0.0236$ \\
Online fraud (e.g., CEO fraud, business email compromise) & $37.24$ & $36.51$ & $0.411$ &  & $0.6817$ \\
Internal threats (e.g., deliberate data deletion, unauthorized data access, unauthorized personal devices) & $36.33$ & $36.76$ & $-0.233$ &  & $0.8156$ \\
\bottomrule
\end{tabular}
\begin{flushleft}
We conducted independent samples $t$ tests to determine whether there are differences in age across groups. Notes: $^{*} p < 0.05, ^{**} p < 0.01, ^{***} p < 0.001$.
\end{flushleft}
\end{table}

Table~\ref{table:age-s} shows the differences in respondents' \textit{age} for groups based on \textit{awareness of solutions}. The results reveal significant differences in average age for awareness of two solutions (i.e., cloud synchronization of data and data backup). Respondents who were aware of these solutions were significantly older than those who were not.

\begin{table}[!htb]
\footnotesize 
\caption{\label{table:age-s} Differences in age across groups based on awareness of solutions.}
\begin{tabular}{lrrrlr}
\toprule
 & Not aware & Aware & $t$ &  & $p$ \\ \midrule
Remote data deletion on lost or stolen devices & $36.17$ & $37.01$ & $-0.502$ &  & $0.6157$ \\
Advanced antimalware solutions (with EDR/XDR capabilities) & $37.50$ & $36.27$ & $0.819$ &  & $0.4133$ \\
Secure connection (e.g., VPN) & $33.92$ & $37.27$ & $-1.800$ &  & $0.0729$ \\
Cloud synchronization of data & $32.73$ & $37.56$ & $-2.570$ & $^{*}$ & $0.0107$ \\
Data backup & $31.96$ & $37.47$ & $-2.940$ & $^{**}$ & $0.0036$ \\
Centralized device management, including mobile device management (MDM) & $37.68$ & $36.41$ & $0.777$ &  & $0.4377$ \\
Advanced firewalls (with IPS/IDS capabilities) & $35.78$ & $36.96$ & $-0.670$ &  & $0.5033$ \\
Training on secure use of devices & $34.42$ & $37.23$ & $-1.616$ &  & $0.1073$ \\
Multi-factor authentication (e.g., 2FA) & $36.35$ & $36.85$ & $-0.302$ &  & $0.7628$ \\
Security operation center (SOC) 24/7 & $38.24$ & $35.58$ & $1.813$ &  & $0.0710$ \\
Centralized management of software updates & $36.07$ & $36.84$ & $-0.473$ &  & $0.6369$ \\
Organizational critical infrastructure access control & $36.75$ & $36.59$ & $0.098$ &  & $0.9219$ \\
\bottomrule
\end{tabular}
\begin{flushleft}
We conducted independent samples $t$ tests to determine whether there are differences in age across groups. Notes: $^{*} p < 0.05, ^{**} p < 0.01, ^{***} p < 0.001$.
\end{flushleft}
\end{table}

These results indicate that decision-makers who are aware of certain threats and solutions are older than decision-makers who are not. However, this is true only for a minority of threats ($33.3\%$) and solutions ($16.7\%$) included in our study.

\subsubsection{Experience with information security}

Table~\ref{table:isec-t} presents the differences in respondents' \textit{experience with information security} for groups based on \textit{awareness of threats}. The results show significant differences in average experience with information security for awareness of four threats (i.e., loss of access to data, DDoS attacks, malware infection and phishing). Respondents who were aware of these threats had significantly more experience with information security than those who were not.

\begin{table}[!htb]
\footnotesize 
\caption{\label{table:isec-t} Differences in experience with information security across groups based on awareness of threats.}
\begin{tabular}{lrrrlr}
\toprule
 & Not aware & Aware & $t$ &  & $p$ \\ \midrule
Loss of access to data (e.g., ransomware, locking of devices) & $7.52$ & $10.50$ & $-2.665$ & $^{**}$ & $0.0092$ \\
Information system intrusion (e.g., hacking) & $8.74$ & $10.13$ & $-0.991$ &  & $0.3225$ \\
Theft of business-critical data (industrial espionage) & $8.37$ & $10.51$ & $-1.968$ &  & $0.0501$ \\
Distributed denial of service (DDoS) attacks & $8.57$ & $10.65$ & $-2.026$ & $^{*}$ & $0.0437$ \\
Takeover of devices (e.g., botnets) & $8.70$ & $10.54$ & $-1.752$ &  & $0.0809$ \\
Malware infection (e.g., viruses, worms, Trojans, spyware) & $7.03$ & $10.25$ & $-2.554$ & $^{*}$ & $0.0137$ \\
Phishing & $8.14$ & $10.50$ & $-2.083$ & $^{*}$ & $0.0382$ \\
Online fraud (e.g., CEO fraud, business email compromise) & $8.07$ & $10.40$ & $-1.940$ &  & $0.0534$ \\
Internal threats (e.g., deliberate data deletion, unauthorized data access, unauthorized personal devices) & $8.48$ & $10.26$ & $-1.437$ &  & $0.1519$ \\
\bottomrule
\end{tabular}
\begin{flushleft}
We conducted independent samples $t$ tests to determine whether there are differences in experience with information security across groups. Notes: $^{*} p < 0.05, ^{**} p < 0.01, ^{***} p < 0.001$.
\end{flushleft}
\end{table}

Table~\ref{table:isec-s} presents the differences in respondents' \textit{experience with information security} for groups based on \textit{awareness of solutions}. The results show significant differences in average experience with information security for awareness of seven solutions (i.e., remote data deletion, advanced antimalware solutions with EDR/XDR capabilities, secure connection, cloud synchronization of data, centralized device management, advanced firewalls and training). Respondents who were aware of these solutions had significantly more experience with information security than those who were not.

\begin{table}[!htb]
\footnotesize 
\caption{\label{table:isec-s} Differences in experience with information security across groups based on awareness of solutions.}
\begin{tabular}{lrrrlr}
\toprule
 & Not aware & Aware & $t$ &  & $p$ \\ \midrule
Remote data deletion on lost or stolen devices & $8.21$ & $10.51$ & $-2.045$ & $^{*}$ & $0.0419$ \\
Advanced antimalware solutions (with EDR/XDR capabilities) & $8.11$ & $10.83$ & $-2.919$ & $^{**}$ & $0.0039$ \\
Secure connection (e.g., VPN) & $6.90$ & $10.60$ & $-3.318$ & $^{**}$ & $0.0014$ \\
Cloud synchronization of data & $7.41$ & $10.44$ & $-2.332$ & $^{*}$ & $0.0204$ \\
Data backup & $8.70$ & $10.05$ & $-1.060$ &  & $0.2901$ \\
Centralized device management, including mobile device management (MDM) & $7.63$ & $10.75$ & $-2.824$ & $^{**}$ & $0.0051$ \\
Advanced firewalls (with IPS/IDS capabilities) & $7.60$ & $10.51$ & $-2.410$ & $^{*}$ & $0.0166$ \\
Training on secure use of devices & $7.20$ & $10.54$ & $-3.205$ & $^{**}$ & $0.0018$ \\
Multi-factor authentication (e.g., 2FA) & $9.22$ & $10.19$ & $-0.889$ &  & $0.3750$ \\
Security operation center (SOC) 24/7 & $9.04$ & $10.30$ & $-1.286$ &  & $0.1994$ \\
Centralized management of software updates & $8.51$ & $10.45$ & $-1.713$ &  & $0.0878$ \\
Organizational critical infrastructure access control & $8.78$ & $10.25$ & $-1.381$ &  & $0.1684$ \\
\bottomrule
\end{tabular}
\begin{flushleft}
We conducted independent samples $t$ tests to determine whether there are differences in experience with information security across groups. Notes: $^{*} p < 0.05, ^{**} p < 0.01, ^{***} p < 0.001$.
\end{flushleft}
\end{table}

These results indicate that decision-makers who are aware of approximately a half of threats ($44.4\%$) and solutions ($58.3\%$) included in our study have more experience with information security than decision-makers who are not.

\subsubsection{Experience with IT}

Table~\ref{table:it-t} shows the differences in respondents' \textit{experience with IT} for groups based on \textit{awareness of threats}. The results indicate significant differences in average experience with IT for awareness of six threats (i.e., loss of access to data, industrial espionage, DDoS attacks, botnets, malware infection and phishing). Respondents who were aware of these threats had significantly more experience with IT than those who were not.

\begin{table}[!htb]
\footnotesize 
\caption{\label{table:it-t} Differences in experience with IT across groups based on awareness of threats.}
\begin{tabular}{lrrrlr}
\toprule
 & Not aware & Aware & $t$ &  & $p$ \\ \midrule
Loss of access to data (e.g., ransomware, locking of devices) & $7.36$ & $11.17$ & $-2.959$ & $^{**}$ & $0.0034$ \\
Information system intrusion (e.g., hacking) & $8.95$ & $10.67$ & $-1.165$ &  & $0.2451$ \\
Theft of business-critical data (industrial espionage) & $7.84$ & $11.38$ & $-3.374$ & $^{***}$ & $0.0009$ \\
Distributed denial of service (DDoS) attacks & $8.89$ & $11.25$ & $-2.189$ & $^{*}$ & $0.0294$ \\
Takeover of devices (e.g., botnets) & $8.51$ & $11.38$ & $-2.618$ & $^{**}$ & $0.0093$ \\
Malware infection (e.g., viruses, worms, Trojans, spyware) & $6.86$ & $10.85$ & $-3.076$ & $^{**}$ & $0.0034$ \\
Phishing & $8.08$ & $11.16$ & $-2.585$ & $^{*}$ & $0.0102$ \\
Online fraud (e.g., CEO fraud, business email compromise) & $8.93$ & $10.81$ & $-1.486$ &  & $0.1385$ \\
Internal threats (e.g., deliberate data deletion, unauthorized data access, unauthorized personal devices) & $8.38$ & $10.83$ & $-1.863$ &  & $0.0636$ \\
\bottomrule
\end{tabular}
\begin{flushleft}
We conducted independent samples $t$ tests to determine whether there are differences in experience with IT across groups. Notes: $^{*} p < 0.05, ^{**} p < 0.01, ^{***} p < 0.001$.
\end{flushleft}
\end{table}

Table~\ref{table:it-s} presents the differences in respondents' \textit{experience with IT} for groups based on \textit{awareness of solutions}. The results show significant differences in average experience with IT for awareness of eight solutions (i.e., remote data deletion, advanced antimalware solutions with EDR/XDR capabilities, secure connection, cloud synchronization of data, data backup, centralized device management, advanced firewalls and training). Respondents who were aware of these solutions had significantly more experience with IT than those who were not.

\begin{table}[!htb]
\footnotesize 
\caption{\label{table:it-s} Differences in experience with IT across groups based on awareness of solutions.}
\begin{tabular}{lrrrlr}
\toprule
 & Not aware & Aware & $t$ &  & $p$ \\ \midrule
Remote data deletion on lost or stolen devices & $8.44$ & $11.10$ & $-2.235$ & $^{*}$ & $0.0262$ \\
Advanced antimalware solutions (with EDR/XDR capabilities) & $8.67$ & $11.30$ & $-2.501$ & $^{*}$ & $0.0130$ \\
Secure connection (e.g., VPN) & $7.08$ & $11.16$ & $-3.106$ & $^{**}$ & $0.0021$ \\
Cloud synchronization of data & $7.39$ & $11.03$ & $-2.664$ & $^{**}$ & $0.0082$ \\
Data backup & $7.83$ & $10.81$ & $-2.219$ & $^{*}$ & $0.0273$ \\
Centralized device management, including mobile device management (MDM) & $7.70$ & $11.39$ & $-3.172$ & $^{**}$ & $0.0017$ \\
Advanced firewalls (with IPS/IDS capabilities) & $7.71$ & $11.11$ & $-2.686$ & $^{**}$ & $0.0077$ \\
Training on secure use of devices & $6.85$ & $11.26$ & $-4.192$ & $^{***}$ & $0.0000$ \\
Multi-factor authentication (e.g., 2FA) & $9.41$ & $10.80$ & $-1.200$ &  & $0.2312$ \\
Security operation center (SOC) 24/7 & $9.60$ & $10.78$ & $-1.138$ &  & $0.2563$ \\
Centralized management of software updates & $8.93$ & $10.94$ & $-1.698$ &  & $0.0907$ \\
Organizational critical infrastructure access control & $9.18$ & $10.79$ & $-1.421$ &  & $0.1564$ \\
\bottomrule
\end{tabular}
\begin{flushleft}
We conducted independent samples $t$ tests to determine whether there are differences in experience with IT across groups. Notes: $^{*} p < 0.05, ^{**} p < 0.01, ^{***} p < 0.001$.
\end{flushleft}
\end{table}

These results indicate that decision-makers who are aware of two thirds of threats ($66.7\%$) and solutions ($66.7\%$) included in our study have more experience with IT than decision-makers who are not.

\section{Discussion}
\label{section:discussion}

In this section, we first provide theoretical and practical implications of this study. Next, we discuss its limitations and put forward directions for future research.

\subsection{Theoretical implications}

This study makes a number of theoretical contributions to the literature. First, awareness of well-known threats and solutions seems to be quite low for cyber and information security decision-makers. The results of this study indicate that about a third of decision-makers are not aware of DDoS attacks ($35.1\%$) and botnets ($32.4\%$), and about a quarter are not aware of industrial espionage ($27.2\%$) and phishing ($24.4\%$). This is especially surprising since these threats are among the currently most prevalent ones \cite{CERT-EU2024,ENISA2023}. Awareness of solutions does not appear to be much better. A quarter or more decision-makers are not aware of seven out of 12 solutions included in our study. More than a third of decision-makers were unaware of two of these solutions, namely SOC ($38.4\%$) and advanced antimalware solutions with EDR/XDR capabilities ($36.2\%$), and a quarter or more decision-makers were unaware of further five solutions (i.e., organizational critical infrastructure access control ($28.2\%$), centralized device management, including mobile device management (MDM) ($26.7\%$), multi-factor authentication (e.g., 2FA) ($26.6\%$), centralized management of software updates ($26.3\%$), remote data deletion on lost or stolen devices ($25.0\%$)). Even though cyber and information security decision-makers are the primary enablers of cybersecurity in organizations, the results of our study suggest they are insufficiently aware of both threats and solutions. This lack of decision-makers' cybersecurity awareness may significantly impact their ability to make adequate decisions regarding cybersecurity, which is a challenging task to start with \cite{Liu2022}. Additionally, this may hinder their ability to lead their organizations towards cyber-resilient culture \cite{Loonam2020,Triplett2022}. These findings are in line with most published literature (e.g., \cite{Drape2021,Rawindaran2022}). This study makes a contribution to the literature on awareness of decision-makers by breaking down cybersecurity awareness into various kinds of threats and solutions. This break-down view shows that cybersecurity awareness may not be a monolithic construct thus future studies may incorporate its different dimensions in their research designs.

Second, this study suggests that there are differences in cybersecurity awareness of decision-makers across groups based on adoption of antimalware solutions in their organizations. The results of our study indicate that awareness of certain but not all threats and solutions is positively associated with either adoption of antimalware solution types or adoption of SOC realizations. These findings contribute to the literature on awareness of decision-makers in organizations adopting different types of advanced antimalware solutions. We need to note that we did not search for causal relationships in our study. Therefore, it remains unclear whether higher awareness of decision-makers in organizations adopting advanced antimalware solutions with EDR/XDR capabilities or organizations adopting either an internal or external SOC is the consequence of this adoption or vice versa. Future studies may thus focus on investigating this causal relationship.

Third, the results of our study indicate that there are differences in cybersecurity awareness of decision-makers across their organizational role types. Non-IT/IS executive decision makers had the least awareness which may be the consequence of their background, and appears to be in line with existing literature \cite{Sapanca2022}. Next, non-executive decision-makers seem to be more aware of certain solutions than other decision-makers which may be a consequence of their more operational involvement in ensuring cybersecurity. This adds to the literature as one of the first studies to investigate the importance of organizational role types. The results of our study however do not support the role of organizational size as suggested by published literature \cite{Bongiovanni2022,Rawindaran2022}. The reason for this divergence could be found in our target population which included decision-makers at both executive and non-executive levels, and with varying backgrounds (i.e., IT/IS and non-IT/IS). Future studies may focus on each type of organizational role type individually.

Fourth, our study identifies several personal characteristics of decision-makers associated with their cybersecurity awareness. Gender seems to be the most important demographic associated with cybersecurity awareness of decision-makers. Similarly to the published literature \cite{Sapanca2022}, male decision-makers were more aware than their female counterparts. Age was also associated with cybersecurity awareness although for a much lower share of threats and solutions. Contrary to the published literature \cite{Rawindaran2022}, formal education was not a significant factor in cybersecurity awareness of decision-makers. Experience with IT was the most seminal personal characteristic associated with cybersecurity awareness of decision-makers, surpassing the share of significant threats and solutions compared to experience with information security. Although this is somewhat surprising, the importance of IT has been emphasized in the literature before \cite{Auyporn2023}. These findings contribute to the literature on the relation between personal characteristics of decision-makers and their cybersecurity awareness. Future works may consider incorporating these characteristics in their research models.

\subsection{Practical implications}

This paper provides some practical implications for increasing the ability to achieve higher levels of cyber-resilience in organizations. Since cyber and information security decision-makers are the main drivers and enablers of the cybersecurity mindset in organizations, understanding and improving their cybersecurity awareness has the potential to further improve the overall cybersecurity of organizations. First, this study provides insights into which threats and solutions are less known among cyber and information security decision-makers. Therefore, it provides straight-forward guidance on which threats and solutions need to be better promoted among cyber and information security decision-makers.

Second, the results of our study identify which decision-makers are more likely to be less aware of threats and solutions -- i.e., especially decision-makers in organizations not adopting a malware solution, decision-makers in organizations not adopting a SOC, non-IT/IS executive decision makers, female and younger decision-makers, and decision-makers with less experience with IT or information security. These characteristics can help to target the most needy decision-makers with cybersecurity awareness interventions. Detailed insights from this study may additionally help to adapt such interventions to the needs of a specific subgroup of decision-makers.

\subsection{Limitations and future research}

This study has some limitations that the readers should note. First, the study was conducted in a single cultural context. Since cultural contexts may be an important factor when studying cybersecurity awareness, the findings of this study may not be generalized to other cultural contexts. Future studies comparing our findings to or investigating the influence of different cultural contexts would help to further generalize the findings of our study. Second, the questionnaire included single items for measured constructs, such as dimensions of threats and solutions. This significantly affects the ability to check reliability and validity of the measurement instrument. Although we addressed this issue by merging scores for both awareness constructs, future studies may be needed to further confirm our findings. Third, we are unsure whether there were cases of more than one respondent from a single organization since we did not ask respondents to name organizations in which they were employed. If there were several such cases, it may partially influence the results related to organizational factors, such as \textit{adopted antimalware solution type}, \textit{adopted SOC realization}, and \textit{organization size}. Even though the possibility of this are relatively low, future studies would be beneficial for confirming the findings of our study.

\begin{acks}
\small
The work of SV and BM was partially funded by A1 Slovenia (\url{https://www.a1.si/}) as part of the \textit{Study on cybersecurity} project, and by Ministry of Higher Education, Science, and Innovation of the Republic of Slovenia and the European Union's NextGenerationEU initiative as part of the \textit{Green and Resilient Transition for a Safe and Successful Society} project. The funders had no role in data analysis, decision to publish, or preparation of the manuscript. A1 Slovenia had a role in study design as they were included in narrowing down the list of potential threats and solutions compiled by the authors. The final decision on the included threats and solution in the study rested with the authors. A1 Slovenia additionally had a role in data collection as a funder and an intermediary between the authors and the CINT platform. Other funders had no role in study design or data collection.
\end{acks}

\bibliographystyle{ACM-Reference-Format}

\begin{thebibliography}{38}


\ifx \showCODEN    \undefined \def \showCODEN     #1{\unskip}     \fi
\ifx \showDOI      \undefined \def \showDOI       #1{#1}\fi
\ifx \showISBNx    \undefined \def \showISBNx     #1{\unskip}     \fi
\ifx \showISBNxiii \undefined \def \showISBNxiii  #1{\unskip}     \fi
\ifx \showISSN     \undefined \def \showISSN      #1{\unskip}     \fi
\ifx \showLCCN     \undefined \def \showLCCN      #1{\unskip}     \fi
\ifx \shownote     \undefined \def \shownote      #1{#1}          \fi
\ifx \showarticletitle \undefined \def \showarticletitle #1{#1}   \fi
\ifx \showURL      \undefined \def \showURL       {\relax}        \fi
\providecommand\bibfield[2]{#2}
\providecommand\bibinfo[2]{#2}
\providecommand\natexlab[1]{#1}
\providecommand\showeprint[2][]{arXiv:#2}

\bibitem[Auyporn et~al\mbox{.}(2023)]%
        {Auyporn2023}
\bibfield{author}{\bibinfo{person}{Wipawadee Auyporn}, \bibinfo{person}{Krerk Piromsopa}, {and} \bibinfo{person}{Thitivadee Chaiyawat}.} \bibinfo{year}{2023}\natexlab{}.
\newblock \showarticletitle{A Study of Distinguishing Factors between SME Adopters versus Non-Adopters of Cybersecurity Standard}.
\newblock \bibinfo{journal}{\emph{International Journal of Computing and Digital Systems}} \bibinfo{volume}{13}, \bibinfo{number}{1} (\bibinfo{year}{2023}), \bibinfo{pages}{189--198}.
\newblock
\urldef\tempurl%
\url{https://doi.org/10.12785/ijcds/130153}
\showDOI{\tempurl}


\bibitem[Batrachenko et~al\mbox{.}(2024)]%
        {Batrachenko2024}
\bibfield{author}{\bibinfo{person}{Tetiana Batrachenko}, \bibinfo{person}{Iryna Lehan}, \bibinfo{person}{Vitalii Kuchmenko}, \bibinfo{person}{Volodymyr Kovalchuk}, {and} \bibinfo{person}{Olha Mazurenko}.} \bibinfo{year}{2024}\natexlab{}.
\newblock \showarticletitle{Cybercrime in the context of the digital age: analysis of threats, legal challenges and strategies}.
\newblock \bibinfo{journal}{\emph{Multidisciplinary Science Journal}}  \bibinfo{volume}{6} (\bibinfo{year}{2024}), \bibinfo{pages}{e2024ss0212}.
\newblock
\urldef\tempurl%
\url{https://doi.org/10.31893/multiscience.2024ss0212}
\showDOI{\tempurl}


\bibitem[Bongiovanni et~al\mbox{.}(2022)]%
        {Bongiovanni2022}
\bibfield{author}{\bibinfo{person}{Ivano Bongiovanni}, \bibinfo{person}{Karen Renaud}, \bibinfo{person}{Humphrey Brydon}, \bibinfo{person}{Renette Blignaut}, {and} \bibinfo{person}{Angelo Cavallo}.} \bibinfo{year}{2022}\natexlab{}.
\newblock \showarticletitle{A quantification mechanism for assessing adherence to information security governance guidelines}.
\newblock \bibinfo{journal}{\emph{Information \& Computer Security}} \bibinfo{volume}{30}, \bibinfo{number}{4} (\bibinfo{year}{2022}), \bibinfo{pages}{517--548}.
\newblock
\urldef\tempurl%
\url{https://doi.org/10.1108/ICS-08-2021-0112}
\showDOI{\tempurl}


\bibitem[Brockinton et~al\mbox{.}(2022)]%
        {Brockinton2022}
\bibfield{author}{\bibinfo{person}{Amanda Brockinton}, \bibinfo{person}{Sam Hirst}, \bibinfo{person}{Ruijie Wang}, \bibinfo{person}{John McAlaney}, {and} \bibinfo{person}{Shelley Thompson}.} \bibinfo{year}{2022}\natexlab{}.
\newblock \showarticletitle{Utilising online eye-tracking to discern the impacts of cultural backgrounds on fake and real news decision-making}.
\newblock \bibinfo{journal}{\emph{Frontiers in Psychology}}  \bibinfo{volume}{13} (\bibinfo{year}{2022}), \bibinfo{pages}{999780}.
\newblock
\urldef\tempurl%
\url{https://doi.org/10.3389/fpsyg.2022.999780}
\showDOI{\tempurl}


\bibitem[{CERT-EU}(2024)]%
        {CERT-EU2024}
\bibfield{author}{\bibinfo{person}{{CERT-EU}}.} \bibinfo{year}{2024}\natexlab{}.
\newblock \bibinfo{booktitle}{\emph{{Threat Landscape Report 2023 - Year Review}}}.
\newblock \bibinfo{type}{{T}echnical {R}eport}. \bibinfo{institution}{{CERT-EU}}. \bibinfo{pages}{1--21} pages.
\newblock


\bibitem[Cochran and Napshin(2021)]%
        {Cochran2021}
\bibfield{author}{\bibinfo{person}{Justin~D Cochran} {and} \bibinfo{person}{Stuart~A Napshin}.} \bibinfo{year}{2021}\natexlab{}.
\newblock \showarticletitle{Deepfakes: awareness, concerns, and platform accountability}.
\newblock \bibinfo{journal}{\emph{Cyberpsychology, Behavior, and Social Networking}} \bibinfo{volume}{24}, \bibinfo{number}{3} (\bibinfo{year}{2021}), \bibinfo{pages}{164--172}.
\newblock
\urldef\tempurl%
\url{https://doi.org/10.1089/cyber.2020.0100}
\showDOI{\tempurl}


\bibitem[Cuchta et~al\mbox{.}(2023)]%
        {Cuchta2023}
\bibfield{author}{\bibinfo{person}{Tom Cuchta}, \bibinfo{person}{Brian Blackwood}, \bibinfo{person}{Thomas~R. Devine}, {and} \bibinfo{person}{Robert~J. Niichel}.} \bibinfo{year}{2023}\natexlab{}.
\newblock \showarticletitle{Human risk factors in cybersecurity: Experimental assessment of an academic human attack surface}.
\newblock \bibinfo{journal}{\emph{Interaction Studies}} \bibinfo{volume}{24}, \bibinfo{number}{3} (\bibinfo{year}{2023}), \bibinfo{pages}{437--463}.
\newblock
\urldef\tempurl%
\url{https://doi.org/10.1075/is.22053.cuc}
\showDOI{\tempurl}


\bibitem[Drape et~al\mbox{.}(2021)]%
        {Drape2021}
\bibfield{author}{\bibinfo{person}{Tiffany Drape}, \bibinfo{person}{Noah Magerkorth}, \bibinfo{person}{Anuradha Sen}, \bibinfo{person}{Joseph Simpson}, \bibinfo{person}{Megan Seibel}, \bibinfo{person}{Randall~Steven Murch}, {and} \bibinfo{person}{Susan~E. Duncan}.} \bibinfo{year}{2021}\natexlab{}.
\newblock \showarticletitle{Assessing the Role of Cyberbiosecurity in Agriculture: A Case Study}.
\newblock \bibinfo{journal}{\emph{Frontiers in Bioengineering and Biotechnology}}  \bibinfo{volume}{9} (\bibinfo{year}{2021}), \bibinfo{pages}{737927}.
\newblock
\urldef\tempurl%
\url{https://doi.org/10.3389/fbioe.2021.737927}
\showDOI{\tempurl}


\bibitem[Dykstra et~al\mbox{.}(2023)]%
        {Dykstra2023}
\bibfield{author}{\bibinfo{person}{Josiah Dykstra}, \bibinfo{person}{Lawrence~A Gordon}, \bibinfo{person}{Martin~P Loeb}, {and} \bibinfo{person}{Lei Zhou}.} \bibinfo{year}{2023}\natexlab{}.
\newblock \showarticletitle{{Maximizing the benefits from sharing cyber threat intelligence by government agencies and departments}}.
\newblock \bibinfo{journal}{\emph{Journal of Cybersecurity}} \bibinfo{volume}{9}, \bibinfo{number}{1} (\bibinfo{year}{2023}), \bibinfo{pages}{tyad003}.
\newblock
\urldef\tempurl%
\url{https://doi.org/10.1093/cybsec/tyad003}
\showDOI{\tempurl}


\bibitem[Ebert et~al\mbox{.}(2023)]%
        {Ebert2023}
\bibfield{author}{\bibinfo{person}{Nico Ebert}, \bibinfo{person}{Thierry Schaltegger}, \bibinfo{person}{Benjamin Ambuehl}, \bibinfo{person}{Lorin Schöni}, \bibinfo{person}{Verena Zimmermann}, {and} \bibinfo{person}{Melanie Knieps}.} \bibinfo{year}{2023}\natexlab{}.
\newblock \showarticletitle{Learning from safety science: A way forward for studying cybersecurity incidents in organizations}.
\newblock \bibinfo{journal}{\emph{Computers \& Security}}  \bibinfo{volume}{134} (\bibinfo{year}{2023}), \bibinfo{pages}{103435}.
\newblock
\urldef\tempurl%
\url{https://doi.org/10.1016/j.cose.2023.103435}
\showDOI{\tempurl}


\bibitem[{ENISA}(2023)]%
        {ENISA2023}
\bibfield{author}{\bibinfo{person}{{ENISA}}.} \bibinfo{year}{2023}\natexlab{}.
\newblock \bibinfo{booktitle}{\emph{{ENISA Threat Landscape 2023}}}.
\newblock \bibinfo{type}{{T}echnical {R}eport}. \bibinfo{institution}{{ENISA}}. \bibinfo{pages}{1--160} pages.
\newblock
\urldef\tempurl%
\url{https://doi.org/10.2824/782573}
\showDOI{\tempurl}


\bibitem[Fujs et~al\mbox{.}(2023)]%
        {Fujs2023:cs}
\bibfield{author}{\bibinfo{person}{Damjan Fujs}, \bibinfo{person}{Simon Vrhovec}, {and} \bibinfo{person}{Damjan Vavpotič}.} \bibinfo{year}{2023}\natexlab{}.
\newblock \showarticletitle{{Balancing software and training requirements for information security}}.
\newblock \bibinfo{journal}{\emph{Computers \& Security}}  \bibinfo{volume}{134} (\bibinfo{year}{2023}), \bibinfo{pages}{103467:1--13}.
\newblock
\urldef\tempurl%
\url{https://doi.org/10.1016/j.cose.2023.103467}
\showDOI{\tempurl}


\bibitem[Gomez and Shandler(2024)]%
        {Gomez2024}
\bibfield{author}{\bibinfo{person}{Miguel~Alberto Gomez} {and} \bibinfo{person}{Ryan Shandler}.} \bibinfo{year}{2024}\natexlab{}.
\newblock \showarticletitle{{Trust at Risk: The Effect of Proximity to Cyberattacks}}.
\newblock \bibinfo{journal}{\emph{Journal of Global Security Studies}} \bibinfo{volume}{9}, \bibinfo{number}{2} (\bibinfo{year}{2024}), \bibinfo{pages}{ogae002}.
\newblock
\urldef\tempurl%
\url{https://doi.org/10.1093/jogss/ogae002}
\showDOI{\tempurl}


\bibitem[Kianpour et~al\mbox{.}(2021)]%
        {Kianpour2021}
\bibfield{author}{\bibinfo{person}{Mazaher Kianpour}, \bibinfo{person}{Stewart~J. Kowalski}, {and} \bibinfo{person}{Harald Øverby}.} \bibinfo{year}{2021}\natexlab{}.
\newblock \showarticletitle{Systematically Understanding Cybersecurity Economics: A Survey}.
\newblock \bibinfo{journal}{\emph{Sustainability}} \bibinfo{volume}{13}, \bibinfo{number}{24} (\bibinfo{year}{2021}), \bibinfo{pages}{136771:1--28}.
\newblock
\urldef\tempurl%
\url{https://doi.org/10.3390/su132413677}
\showDOI{\tempurl}


\bibitem[Lenz et~al\mbox{.}(2023)]%
        {Lenz2023:cs}
\bibfield{author}{\bibinfo{person}{Julia Lenz}, \bibinfo{person}{Zdravko Bozakov}, \bibinfo{person}{Steffen Wendzel}, {and} \bibinfo{person}{Simon Vrhovec}.} \bibinfo{year}{2023}\natexlab{}.
\newblock \showarticletitle{{Why People Replace their Aging Smart Devices: A Push--Pull--Mooring Perspective}}.
\newblock \bibinfo{journal}{\emph{Computers \& Security}}  \bibinfo{volume}{130} (\bibinfo{year}{2023}), \bibinfo{pages}{103258:1--22}.
\newblock
\urldef\tempurl%
\url{https://doi.org/10.1016/j.cose.2023.103258}
\showDOI{\tempurl}


\bibitem[Lif et~al\mbox{.}(2022)]%
        {Lif2022}
\bibfield{author}{\bibinfo{person}{Patrik Lif}, \bibinfo{person}{Teodor Sommestad}, \bibinfo{person}{P{\"a}r-Anders Albinsson}, \bibinfo{person}{Christian Valassi}, {and} \bibinfo{person}{Daniel Eidenskog}.} \bibinfo{year}{2022}\natexlab{}.
\newblock \showarticletitle{Validation of Cyber Test for Future Soldiers: A Test Battery for the Selection of Cyber Soldiers}.
\newblock \bibinfo{journal}{\emph{Frontiers in Psychology}}  \bibinfo{volume}{13} (\bibinfo{year}{2022}), \bibinfo{pages}{868311}.
\newblock
\urldef\tempurl%
\url{https://doi.org/10.3389/fpsyg.2022.868311}
\showDOI{\tempurl}


\bibitem[Liu et~al\mbox{.}(2022)]%
        {Liu2022}
\bibfield{author}{\bibinfo{person}{Xiang Liu}, \bibinfo{person}{Sayed~Fayaz Ahmad}, \bibinfo{person}{Muhammad~Khalid Anser}, \bibinfo{person}{Jingying Ke}, \bibinfo{person}{Muhammad Irshad}, \bibinfo{person}{Jabbar Ul-Haq}, {and} \bibinfo{person}{Shujaat Abbas}.} \bibinfo{year}{2022}\natexlab{}.
\newblock \showarticletitle{Cyber security threats: A never-ending challenge for e-commerce}.
\newblock \bibinfo{journal}{\emph{Frontiers in psychology}}  \bibinfo{volume}{13} (\bibinfo{year}{2022}), \bibinfo{pages}{927398}.
\newblock
\urldef\tempurl%
\url{https://doi.org/10.3389/fpsyg.2022.927398}
\showDOI{\tempurl}


\bibitem[Loonam et~al\mbox{.}(2020)]%
        {Loonam2020}
\bibfield{author}{\bibinfo{person}{John Loonam}, \bibinfo{person}{Jeremy Zwiegelaar}, \bibinfo{person}{Vikas Kumar}, {and} \bibinfo{person}{Charles Booth}.} \bibinfo{year}{2020}\natexlab{}.
\newblock \showarticletitle{Cyber-resiliency for digital enterprises: a strategic leadership perspective}.
\newblock \bibinfo{journal}{\emph{IEEE Transactions on Engineering Management}} \bibinfo{volume}{69}, \bibinfo{number}{6} (\bibinfo{year}{2020}), \bibinfo{pages}{3757--3770}.
\newblock
\urldef\tempurl%
\url{https://doi.org/10.1109/TEM.2020.2996175}
\showDOI{\tempurl}


\bibitem[Mikuletič et~al\mbox{.}(2024)]%
        {Mikuletic2024}
\bibfield{author}{\bibinfo{person}{Samanta Mikuletič}, \bibinfo{person}{Simon Vrhovec}, \bibinfo{person}{Brigita Skela-Savić}, {and} \bibinfo{person}{Boštjan Žvanut}.} \bibinfo{year}{2024}\natexlab{}.
\newblock \showarticletitle{{Security and privacy oriented information security culture (ISC): Explaining unauthorized access to healthcare data by nursing employees}}.
\newblock \bibinfo{journal}{\emph{Computers \& Security}}  \bibinfo{volume}{136} (\bibinfo{year}{2024}), \bibinfo{pages}{103489:1--14}.
\newblock
\urldef\tempurl%
\url{https://doi.org/10.1016/j.cose.2023.103489}
\showDOI{\tempurl}


\bibitem[Moyo and Loock(2021)]%
        {Moyo2021}
\bibfield{author}{\bibinfo{person}{Moses Moyo} {and} \bibinfo{person}{Marianne Loock}.} \bibinfo{year}{2021}\natexlab{}.
\newblock \showarticletitle{Conceptualising a Cloud Business Intelligence Security Evaluation Framework for Small and Medium Enterprises in Small Towns of the Limpopo Province, South Africa}.
\newblock \bibinfo{journal}{\emph{Information}} \bibinfo{volume}{12}, \bibinfo{number}{3} (\bibinfo{year}{2021}), \bibinfo{pages}{128:1--27}.
\newblock
\urldef\tempurl%
\url{https://doi.org/10.3390/info12030128}
\showDOI{\tempurl}


\bibitem[Naseer et~al\mbox{.}(2023)]%
        {Naseer2023}
\bibfield{author}{\bibinfo{person}{Ayesha Naseer}, \bibinfo{person}{Humza Naseer}, \bibinfo{person}{Atif Ahmad}, \bibinfo{person}{Sean~B Maynard}, {and} \bibinfo{person}{Adil~Masood Siddiqui}.} \bibinfo{year}{2023}\natexlab{}.
\newblock \showarticletitle{Moving towards agile cybersecurity incident response: A case study exploring the enabling role of big data analytics-embedded dynamic capabilities}.
\newblock \bibinfo{journal}{\emph{Computers \& Security}}  \bibinfo{volume}{135} (\bibinfo{year}{2023}), \bibinfo{pages}{103525}.
\newblock
\urldef\tempurl%
\url{https://doi.org/10.1016/j.cose.2023.103525}
\showDOI{\tempurl}


\bibitem[Parkin et~al\mbox{.}(2023)]%
        {Parkin2023}
\bibfield{author}{\bibinfo{person}{Simon Parkin}, \bibinfo{person}{Kristen Kuhn}, {and} \bibinfo{person}{Siraj~A Shaikh}.} \bibinfo{year}{2023}\natexlab{}.
\newblock \showarticletitle{{Executive decision-makers: a scenario-based approach to assessing organizational cyber-risk perception}}.
\newblock \bibinfo{journal}{\emph{Journal of Cybersecurity}} \bibinfo{volume}{9}, \bibinfo{number}{1} (\bibinfo{year}{2023}), \bibinfo{pages}{tyad018}.
\newblock
\urldef\tempurl%
\url{https://doi.org/10.1093/cybsec/tyad018}
\showDOI{\tempurl}


\bibitem[Piazza et~al\mbox{.}(2023)]%
        {Piazza2023}
\bibfield{author}{\bibinfo{person}{Anna Piazza}, \bibinfo{person}{Srinidhi Vasudevan}, {and} \bibinfo{person}{Madeline Carr}.} \bibinfo{year}{2023}\natexlab{}.
\newblock \showarticletitle{{Cybersecurity in UK Universities: mapping (or managing) threat intelligence sharing within the higher education sector}}.
\newblock \bibinfo{journal}{\emph{Journal of Cybersecurity}} \bibinfo{volume}{9}, \bibinfo{number}{1} (\bibinfo{year}{2023}), \bibinfo{pages}{tyad019}.
\newblock
\urldef\tempurl%
\url{https://doi.org/10.1093/cybsec/tyad019}
\showDOI{\tempurl}


\bibitem[Prebot et~al\mbox{.}(2023)]%
        {Prebot2023}
\bibfield{author}{\bibinfo{person}{Baptiste Prebot}, \bibinfo{person}{Yinuo Du}, {and} \bibinfo{person}{Cleotilde Gonzalez}.} \bibinfo{year}{2023}\natexlab{}.
\newblock \showarticletitle{{Learning about simulated adversaries from human defenders using interactive cyber-defense games}}.
\newblock \bibinfo{journal}{\emph{Journal of Cybersecurity}} \bibinfo{volume}{9}, \bibinfo{number}{1} (\bibinfo{year}{2023}), \bibinfo{pages}{tyad022}.
\newblock
\urldef\tempurl%
\url{https://doi.org/10.1093/cybsec/tyad022}
\showDOI{\tempurl}


\bibitem[Preuveneers and Joosen(2023)]%
        {Preuveneers2023}
\bibfield{author}{\bibinfo{person}{Davy Preuveneers} {and} \bibinfo{person}{Wouter Joosen}.} \bibinfo{year}{2023}\natexlab{}.
\newblock \showarticletitle{Privacy-preserving correlation of cross-organizational cyber threat intelligence with private graph intersections}.
\newblock \bibinfo{journal}{\emph{Computers \& Security}}  \bibinfo{volume}{135} (\bibinfo{year}{2023}), \bibinfo{pages}{103505}.
\newblock
\urldef\tempurl%
\url{https://doi.org/10.1016/j.cose.2023.103505}
\showDOI{\tempurl}


\bibitem[Rawindaran et~al\mbox{.}(2022)]%
        {Rawindaran2022}
\bibfield{author}{\bibinfo{person}{Nisha Rawindaran}, \bibinfo{person}{Ambikesh Jayal}, {and} \bibinfo{person}{Edmond Prakash}.} \bibinfo{year}{2022}\natexlab{}.
\newblock \showarticletitle{Exploration of the impact of cybersecurity awareness on small and medium enterprises (SMEs) in Wales using intelligent software to combat cybercrime}.
\newblock \bibinfo{journal}{\emph{Computers}} \bibinfo{volume}{11}, \bibinfo{number}{12} (\bibinfo{year}{2022}), \bibinfo{pages}{174}.
\newblock
\urldef\tempurl%
\url{https://doi.org/10.3390/computers11120174}
\showDOI{\tempurl}


\bibitem[Reeves and Ashenden(2023)]%
        {Reeves2023}
\bibfield{author}{\bibinfo{person}{Andrew Reeves} {and} \bibinfo{person}{Debi Ashenden}.} \bibinfo{year}{2023}\natexlab{}.
\newblock \showarticletitle{Understanding decision making in security operations centres: building the case for cyber deception technology}.
\newblock \bibinfo{journal}{\emph{Frontiers in Psychology}}  \bibinfo{volume}{14} (\bibinfo{year}{2023}), \bibinfo{pages}{1165705}.
\newblock
\urldef\tempurl%
\url{https://doi.org/10.3389/fpsyg.2023.1165705}
\showDOI{\tempurl}


\bibitem[Roman et~al\mbox{.}(2023)]%
        {Roman2023}
\bibfield{author}{\bibinfo{person}{Rodrigo Roman}, \bibinfo{person}{Cristina Alcaraz}, \bibinfo{person}{Javier Lopez}, {and} \bibinfo{person}{Kouichi Sakurai}.} \bibinfo{year}{2023}\natexlab{}.
\newblock \showarticletitle{Current Perspectives on Securing Critical Infrastructures’ Supply Chains}.
\newblock \bibinfo{journal}{\emph{IEEE Security \& Privacy}} \bibinfo{volume}{21}, \bibinfo{number}{4} (\bibinfo{year}{2023}), \bibinfo{pages}{29--38}.
\newblock
\urldef\tempurl%
\url{https://doi.org/10.1109/MSEC.2023.3247946}
\showDOI{\tempurl}


\bibitem[Sapanca and Kanbul(2022)]%
        {Sapanca2022}
\bibfield{author}{\bibinfo{person}{Hamza~Fatih Sapanca} {and} \bibinfo{person}{Sezer Kanbul}.} \bibinfo{year}{2022}\natexlab{}.
\newblock \showarticletitle{Risk management in digitalized educational environments: Teachers’ information security awareness levels}.
\newblock \bibinfo{journal}{\emph{Frontiers in Psychology}}  \bibinfo{volume}{13} (\bibinfo{year}{2022}), \bibinfo{pages}{986561}.
\newblock
\urldef\tempurl%
\url{https://doi.org/10.3389/fpsyg.2022.986561}
\showDOI{\tempurl}


\bibitem[Smmarwar et~al\mbox{.}(2024)]%
        {Smmarwar2024}
\bibfield{author}{\bibinfo{person}{Santosh~K. Smmarwar}, \bibinfo{person}{Govind~P. Gupta}, {and} \bibinfo{person}{Sanjay Kumar}.} \bibinfo{year}{2024}\natexlab{}.
\newblock \showarticletitle{Android malware detection and identification frameworks by leveraging the machine and deep learning techniques: A comprehensive review}.
\newblock \bibinfo{journal}{\emph{Telematics and Informatics Reports}}  \bibinfo{volume}{14} (\bibinfo{year}{2024}), \bibinfo{pages}{100130}.
\newblock
\urldef\tempurl%
\url{https://doi.org/10.1016/j.teler.2024.100130}
\showDOI{\tempurl}


\bibitem[Song and Ma(2024)]%
        {Song2024}
\bibfield{author}{\bibinfo{person}{Xiedong Song} {and} \bibinfo{person}{Qinmin Ma}.} \bibinfo{year}{2024}\natexlab{}.
\newblock \showarticletitle{Intrusion detection using federated attention neural network for edge enabled internet of things}.
\newblock \bibinfo{journal}{\emph{Journal of Grid Computing}} \bibinfo{volume}{22}, \bibinfo{number}{1} (\bibinfo{year}{2024}), \bibinfo{pages}{1--17}.
\newblock
\urldef\tempurl%
\url{https://doi.org/10.1007/s10723-023-09725-3}
\showDOI{\tempurl}


\bibitem[Thornton-Trump(2023)]%
        {ThorntonTrump2023}
\bibfield{author}{\bibinfo{person}{Ian Thornton-Trump}.} \bibinfo{year}{2023}\natexlab{}.
\newblock \showarticletitle{GOOD, BETTER \& THE BEST SECURITY}.
\newblock \bibinfo{journal}{\emph{EDPACS}} \bibinfo{volume}{68}, \bibinfo{number}{2} (\bibinfo{year}{2023}), \bibinfo{pages}{21--27}.
\newblock
\urldef\tempurl%
\url{https://doi.org/10.1080/07366981.2023.2210009}
\showDOI{\tempurl}


\bibitem[Tian et~al\mbox{.}(2023)]%
        {Tian2023}
\bibfield{author}{\bibinfo{person}{Chuan~(Annie) Tian}, \bibinfo{person}{Matthew~L. Jensen}, {and} \bibinfo{person}{Alexandra Durcikova}.} \bibinfo{year}{2023}\natexlab{}.
\newblock \showarticletitle{Phishing susceptibility across industries: The differential impact of influence techniques}.
\newblock \bibinfo{journal}{\emph{Computers \& Security}}  \bibinfo{volume}{135} (\bibinfo{year}{2023}), \bibinfo{pages}{103487}.
\newblock
\urldef\tempurl%
\url{https://doi.org/10.1016/j.cose.2023.103487}
\showDOI{\tempurl}


\bibitem[Triplett(2022)]%
        {Triplett2022}
\bibfield{author}{\bibinfo{person}{William~J. Triplett}.} \bibinfo{year}{2022}\natexlab{}.
\newblock \showarticletitle{Addressing Human Factors in Cybersecurity Leadership}.
\newblock \bibinfo{journal}{\emph{Journal of Cybersecurity and Privacy}} \bibinfo{volume}{2}, \bibinfo{number}{3} (\bibinfo{year}{2022}), \bibinfo{pages}{573--586}.
\newblock
\urldef\tempurl%
\url{https://doi.org/10.3390/jcp2030029}
\showDOI{\tempurl}


\bibitem[Turner et~al\mbox{.}(2023)]%
        {Turner2023}
\bibfield{author}{\bibinfo{person}{Adam~Brian Turner}, \bibinfo{person}{Stephen McCombie}, {and} \bibinfo{person}{Allon~J. Uhlmann}.} \bibinfo{year}{2023}\natexlab{}.
\newblock \showarticletitle{Ransomware-Bitcoin Threat Intelligence Sharing Using Structured Threat Information Expression}.
\newblock \bibinfo{journal}{\emph{IEEE Security \& Privacy}} \bibinfo{volume}{21}, \bibinfo{number}{3} (\bibinfo{year}{2023}), \bibinfo{pages}{47--57}.
\newblock
\urldef\tempurl%
\url{https://doi.org/10.1109/MSEC.2022.3166282}
\showDOI{\tempurl}


\bibitem[Vrhovec et~al\mbox{.}(2023)]%
        {Vrhovec2023:cs}
\bibfield{author}{\bibinfo{person}{Simon Vrhovec}, \bibinfo{person}{Igor Bernik}, {and} \bibinfo{person}{Blaž Markelj}.} \bibinfo{year}{2023}\natexlab{}.
\newblock \showarticletitle{{Explaining information seeking intentions: Insights from a Slovenian social engineering awareness campaign}}.
\newblock \bibinfo{journal}{\emph{Computers \& Security}}  \bibinfo{volume}{125} (\bibinfo{year}{2023}), \bibinfo{pages}{103038:1--12}.
\newblock
\urldef\tempurl%
\url{https://doi.org/10.1016/j.cose.2022.103038}
\showDOI{\tempurl}


\bibitem[Yin et~al\mbox{.}(2023)]%
        {Yin2023}
\bibfield{author}{\bibinfo{person}{Ya Yin}, \bibinfo{person}{Carol Hsu}, {and} \bibinfo{person}{Zhongyun Zhou}.} \bibinfo{year}{2023}\natexlab{}.
\newblock \showarticletitle{Employees' in-role and extra-role information security behaviors from the P-E fit perspective}.
\newblock \bibinfo{journal}{\emph{Computers \& Security}}  \bibinfo{volume}{133} (\bibinfo{year}{2023}), \bibinfo{pages}{103390}.
\newblock
\urldef\tempurl%
\url{https://doi.org/10.1016/j.cose.2023.103390}
\showDOI{\tempurl}


\bibitem[{Žvanut, Boštjan and Mihelič, Anže}(2024)]%
        {Zvanut2024}
\bibfield{author}{\bibinfo{person}{{Žvanut, Boštjan and Mihelič, Anže}}.} \bibinfo{year}{2024}\natexlab{}.
\newblock \showarticletitle{Qualitative study on domestic social robot adoption and associated security concerns among older adults in Slovenia}.
\newblock \bibinfo{journal}{\emph{Frontiers in Psychology}}  \bibinfo{volume}{15} (\bibinfo{year}{2024}), \bibinfo{pages}{1343077}.
\newblock
\urldef\tempurl%
\url{https://doi.org/10.3389/fpsyg.2024.1343077}
\showDOI{\tempurl}


\end{thebibliography}


\end{document}